\documentclass[aps,pra,twoside,twocolumn,a4paper,reprint,noeprint]{revtex4-2}
\usepackage[utf8]{inputenc}
\usepackage{graphicx}
\usepackage{amsmath}
\usepackage{amsthm}
\usepackage{amsfonts}
\usepackage{amssymb}
\usepackage{mathtools}
\usepackage{bbm}
\usepackage[usenames,dvipsnames,svgnames]{xcolor}
\usepackage[USenglish]{babel}
\usepackage{hyperref}
\usepackage{grffile}
\usepackage{cleveref}
\usepackage{nicematrix}
\Crefname{figure}{Fig.}{Figs.}
\usepackage{multirow}

\usepackage{array}

\newcommand{\boldone}{\mathbbm{1}}

\allowdisplaybreaks
\begin{document}

\title{Manipulation of Majorana wave packets at surfaces\\
of nodal noncentrosymmetric superconductors}

\author{Clara J. Lapp}
\email{clara{\_}johanna.lapp@tu-dresden.de}
\affiliation{Institute of Theoretical Physics, TU Dresden, 01062 Dresden, Germany}
\affiliation{W\"urzburg--Dresden Cluster of Excellence ct.qmat, TU Dresden, 01062 Dresden, Germany}

\author{Julia M. Link}
\email{julia.link@tu-dresden.de}
\affiliation{Institute of Theoretical Physics, TU Dresden, 01062 Dresden, Germany}
\affiliation{W\"urzburg--Dresden Cluster of Excellence ct.qmat, TU Dresden, 01062 Dresden, Germany}

\author{Carsten Timm}
\email{carsten.timm@tu-dresden.de}
\affiliation{Institute of Theoretical Physics, TU Dresden, 01062 Dresden, Germany}
\affiliation{W\"urzburg--Dresden Cluster of Excellence ct.qmat, TU Dresden, 01062 Dresden, Germany}

\begin{abstract}
Nodal noncentrosymmetric superconductors can host zero-energy flat bands of Majorana surface states within the projection of the nodal lines onto the surface Brillouin zone. Thus, these systems can have stationary, localized Majorana wave packets on certain surfaces, which may be a promising platform for quantum computation. Such applications require protocols to manipulate the wave packets in order to move them without destroying their localization or coherence. As a step in this direction, we explore the idea that the surface states have a nontrivial spin polarization, which can couple for example to the magnetization of a ferromagnetic insulator in contact to the surface, via an exchange term in the Hamiltonian. Such a coupling can make the previously flat bands weakly dispersive. We aim to model the motion of spatially localized wave packets under the influence of an exchange field which is changed adiabatically. We calculate the time-evolved wave packet for a model system and discuss which factors influence the direction of motion and the broadening of the wave packet.
\end{abstract}

\maketitle

\section{Introduction}
\label{sec:intro}

It has been shown \cite{TMYYS10, YSTY11, STYY11, BST11, SR11, SBT12, HQS13, SB15, BST11} that time-reversal-symmetric noncentrosymmetric superconductors (NCSs) can support flat bands of zero-energy surface states in part of their surface Brillouin zone (sBZ). In these systems, nodal lines of the bulk gap can occur if the spin-triplet pairing is sufficiently strong compared to the spin-singlet pairing. One can define a winding number that depends on the surface momentum and that is nonzero within the projection of the bulk nodal lines onto the sBZ, protecting zero-energy Majorana surface modes in these regions of the sBZ \cite{TMYYS10, YSTY11, STYY11, BST11, SR11, SBT12, HQS13, SB15, BST11}. These zero-energy surface states are necessarily in the strong-coupling regime \cite{PoL14, CPF15, RRT20} because their kinetic energy vanishes. Moreover, these Majorana modes are potentially interesting for quantum computation \cite{K03, NSSFS08, SLTD10, LSD10, ORO10, BTS13, EF15, SFN15, OO20, RRT20, LT22, LLT24}, which, however, requires braiding to realize quantum gates, i.e., it is necessary to move wave packets around one another while still keeping them well localized and without destroying their coherence. Moving wave packets requires control of the dispersion of the surface modes; we have to make the bands weakly dispersive with band velocity in the direction in which we want to move the wave packets. Clearly, we also have to be able to change the magnitude and direction of this velocity. The application of an electric field does not work because the Majorana modes in momentum space are charge neutral on average. However, they do carry a momentum-dependent spin polarization so that an applied magnetic field or an exchange field leads to a nontrivial dispersion \cite{YSTY11, MCSR13, STB13, BTS13, WLLL13, BST15, LT22}. We here investigate the idea of introducing an exchange field at the surface of the superconductor, which can for example be achieved by bringing it into contact with a ferromagnetic insulator (FI). The exchange field is assumed to change adiabatically. In this paper, we examine the resulting motion of the wave packets. In order to calculate the time evolution of a spatially localized wave packet under the influence of such an exchange field, we employ a transfer-matrix method to calculate the surface states and energies.

The remainder of this paper is organized as follows: In Sec.~\ref{sec:Model system}, we introduce our model Hamiltonian, discuss its symmetries, and give a brief review on the winding number that protects the zero-energy surface states. In Sec.~\ref{sec:Construction of a wave packet}, we derive an analytical method for the calculation of the surface states both in the presence and in the absence of an exchange field. We also discuss the mathematical construction of a wave packet and its time evolution. In Sec.~\ref{sec:C4v results}, we present the numerical results for the time evolution for a system with the point group $C_{4v}$ and discuss which factors have to be considered to construct wave packets that avoid quick delocalization. Finally, we give a short summary and draw conclusions in Sec.~\ref{sec:summary}.

\section{Model system}
\label{sec:Model system}

In this section, we introduce the model considered in this paper and briefly discuss how its symmetries lead to a surface-momentum-dependent winding number that can protect flat bands of zero-energy surface states. We start with the BCS Hamiltonian of a three-dimensional noncentrosymmetric single-band superconductor,
\begin{equation}\label{eq:hamiltonian}
H_{\text{BCS}} = \frac{1}{2} \sum_{\mathbf{k}} \Psi_{\mathbf{k}}^\dagger \mathcal{H}(\mathbf{k})\Psi_{\mathbf{k}},
\end{equation}
which contains the spinors $\Psi_{\mathbf{k}}=(c_{\mathbf{k},\uparrow},c_{\mathbf{k},\downarrow}, c_{-\mathbf{k},\uparrow}^\dagger,c_{-\mathbf{k},\downarrow}^\dagger)^T$ of creation operators $c^\dagger_{\mathbf{k}\sigma}$ and annihilation operators $c_{\mathbf{k}\sigma}$ for electrons of momentum $\mathbf{k}$ and spin $\sigma\in\lbrace\uparrow,\downarrow\rbrace$ and the Bogoliubov--de Gennes (BdG) Hamiltonian
\begin{equation}\label{eq:bdg_hamiltonian}
\mathcal{H}(\mathbf{k}) = \begin{pmatrix}
h_\text{NCS}(\mathbf{k})& \Delta(\mathbf{k})\\
\Delta^\dagger(\mathbf{k})&-h_\text{NCS}^T(-\mathbf{k})
\end{pmatrix}.
\end{equation}
This $4 \times 4$ matrix consists of four blocks, where the diagonal blocks are given by the normal-state Hamiltonian
\begin{equation}\label{eq:normal_state_hamiltonian}
h_\text{NCS}(\mathbf{k})=\epsilon_{\mathbf{k}} \sigma_0 + \mathbf{g}_{\mathbf{k}} \cdot \boldsymbol{\sigma}
\end{equation}
and the off-diagonal blocks by the gap matrix
\begin{equation}\label{eq:gap_matrix}
\Delta(\mathbf{k})=(\Delta^s_{\mathbf{k}} \sigma_0 + \mathbf{d}_{\mathbf{k}} \cdot \boldsymbol{\sigma})\, i \sigma_y.
\end{equation}
The normal-state Hamiltonian consists of a spin-independent dispersion $\epsilon_{\mathbf{k}}$ multiplied by the $2\times 2$ unit matrix $\sigma_0$ and a spin-orbit-coupling (SOC) term $\mathbf{g}_{\mathbf{k}} \cdot \boldsymbol{\sigma}$, where $\mathbf{g}_{\mathbf{k}}=\lambda \mathbf{l_k}$ is the SOC vector with SOC strength $\lambda$ and $\boldsymbol{\sigma}$ is the vector of Pauli matrices. In order to ensure time-reversal symmetry (TRS), the dispersion $\epsilon_{\mathbf{k}}$ has to be even in $\mathbf{k}$, while the SOC vector $\mathbf{g}_{\mathbf{k}}$ has to be odd. The normal-state Hamiltonian can be diagonalized by the eigenvectors
\begin{equation}\label{eq:eigvec_normalstatehamiltonian}
\mathbf{u}^\pm_{\mathbf{k}} = \begin{pmatrix}
  1 \\
  \displaystyle \pm\frac{l^x_{\mathbf{k}}+ i l^y_{\mathbf{k}}}{l^z_{\mathbf{k}}
    \pm |\mathbf{l_k}|}
\end{pmatrix}
\end{equation}
corresponding to eigenvalues $\xi^\pm_{\mathbf{k}} = \epsilon_{\mathbf{k}} \pm |\mathbf{l_k}|$, which form the positive-helicity and negative-helicity bands, respectively.  Here, we denote the components of the SOC vector by $l^i_{\mathbf{k}}$, i.e., $\mathbf{l_k}=(l^x_{\mathbf{k}},l^y_{\mathbf{k}}, l^z_{\mathbf{k}})^T$.

Due to the absence of inversion symmetry, parity is not a good quantum number so that the gap matrix $\Delta(\mathbf{k})$ generically contains both a spin-singlet part $\Delta^s_{\mathbf{k}} \sigma_0$, which is even in $\mathbf{k}$, and a spin-triplet part $\mathbf{d}_{\mathbf{k}} \cdot \boldsymbol{\sigma}$, where $\mathbf{d_k}$ is the spin-triplet-pairing vector, which is odd in $\mathbf{k}$. The vector $\mathbf{d_k}$ is assumed to be parallel to the SOC vector $\mathbf{g_k}$ as this alignment maximizes the critical temperature \cite{BST11, FAKS04}. We thus write $\mathbf{d_k}=\Delta^t_{\mathbf{k}} \mathbf{l_k}$. We here consider the simplest symmetry-allowed order parameters with $\Delta^s_{\mathbf{k}}=\Delta^s=\text{const}$ and $\Delta^t_{\mathbf{k}}=\Delta^t=\text{const}$, i.e., $(s+p)$-wave pairing. As the SOC vector and the spin-triplet-pairing vector are taken to be parallel the eigenvectors $\mathbf{v_k}$ of the gap matrix can be written in terms of the eigenvectors of the normal-state Hamiltonian as $\mathbf{v_k}^\pm = -i\sigma_y \mathbf{u_k}^\pm$, corresponding to the eigenvalues $\Delta^\pm_{\mathbf{k}}=\Delta^s\pm \Delta^t |\mathbf{l_k}|$, which denote the gap on the positive-helicity and the negative-helicity Fermi surfaces.

The specific form of the SOC vector is restricted by the crystallographic point group, which constrains the BdG Hamiltonian according to
\begin{equation}
U_{\tilde{R}} \mathcal{H}(R^{-1}\mathbf{k}) U_{\tilde{R}}^\dagger= \mathcal{H}(\mathbf{k}),
\end{equation}
where $R$ is an orthogonal $3\times 3$ matrix which represents a symmetry in the point group, the matrix $\tilde{R}$ is given by $\tilde{R} = R/\det(R) = \det(R) R$, and $U_{\tilde{R}} = \text{diag}(u_{\tilde{R}},u_{\tilde{R}}^\ast)$, where $u_{\tilde{R}}=\exp [-i\theta(\mathbf{n} \cdot \boldsymbol{\sigma})/2]$, is the spinor representation of $\tilde{R}$. This leads to the constraints
\begin{align}
\mathbf{l}_{\mathbf{k}} &= \tilde{R}\, \mathbf{l}_{R^{-1}\mathbf{k}}, \label{eq:group_relation_l}\\
\epsilon_{\mathbf{k}} &= \epsilon_{R^{-1} \mathbf{k}}
\end{align}
for all $R$ in the point group.

According to the classification of topological insulators and superconductors known as the tenfold way \cite{Z96, AZ97, SRF08, SRF09, Kit09}, a fully gapped Hamiltonian of the form described above belongs to the class DIII. Particle-hole symmetry (PHS) described by
\begin{equation}\label{eq:PHS}
U_C \mathcal{H}(-\mathbf{k})^T U_C^\dagger = -\mathcal{H}(\mathbf{k}) ,
\end{equation}
with $\mathcal{C}=\mathcal{K} U_C$, $\mathcal{K}$ being the complex-conjugation operator, and $U_C=\sigma_x \otimes \sigma_0$ is ensured by the construction of the BdG Hamiltonian and squares to $+\boldone$. TRS $\mathcal{T}=\mathcal{K} U_T$ with $U_T=\sigma_0 \otimes i \sigma_y $ is also present and squares to $-\boldone$ since the normal-state dispersion $\epsilon_{\mathbf{k}}$ is even in $\mathbf{k}$, while the SOC vector, which couples to the Pauli matrices, is odd, which leads to
\begin{equation}
U_T \mathcal{H}(-\mathbf{k})^T U_T^\dagger = +\mathcal{H}(\mathbf{k}).
\end{equation}
The combination of these antiunitary symmetries is the chiral symmetry
\begin{equation}
U_S^\dagger \mathcal{H}(\mathbf{k}) U_S = -\mathcal{H}(\mathbf{k}),
\end{equation}
i.e., the BdG Hamiltonian anticommutes with the unitary matrix $U_S=i U_T U_C=-\sigma_x \otimes \sigma_y$, which is the essential ingredient leading to a surface-momentum-dependent winding number and zero-energy surface states: If we perform a transformation with a unitary matrix $W_S$ that diagonalizes $U_S$, $W_S U_S W_S^\dagger = \sigma_z \otimes \sigma_0$, the Hamiltonian becomes block-off-diagonal, i.e.,
\begin{equation}\label{eq:H_offdiagonal}
\tilde{\mathcal{H}}(\mathbf{k}) = W_S \mathcal{H}(\mathbf{k}) W_S^\dagger=\begin{pmatrix}
  0 & D(\mathbf{k}) \\ D^\dagger(\mathbf{k}) & 0
\end{pmatrix}.
\end{equation}
For sufficiently large triplet-to-singlet pairing ratio $\Delta^t/\Delta^s$, the gap $\Delta^-_{\mathbf{k}}$ has nodal lines, i.e., one-dimensional regions on the negative-helicity Fermi surface, $\xi^-_{\mathbf{k}}=0$, where the gap $\Delta^-_{\mathbf{k}}$ vanishes. Hence, no global topological invariant can be defined. In the following, we will instead define a winding number that depends on the momentum in the sBZ.

For any given surface orientation, we distinguish between the momentum components $\mathbf{k}_\|$ parallel to the surface and the perpendicular component $k_\perp$. By holding $\mathbf{k}_\|$ fixed and treating only $k_\perp$ as a momentum argument, we obtain the Hamiltonian $\mathcal{H}_{\mathbf{k}_\|}(k_\perp)$ of an effectively one-dimensional system.
For general $\mathbf{k}_\|$, this one-dimensional system does not have TRS or PHS anymore but retains chiral symmetry, which does not change $\mathbf{k}_\|$. The off-diagonal block $D(\mathbf{k})$ in Eq.~\eqref{eq:H_offdiagonal} can now be used to define a $\mathbf{k}_\|$-dependent winding number
\begin{equation}
w(\mathbf{k}_\|) = \frac{1}{2\pi} \int_{k_\perp} dk_\perp\, \partial_{k_\perp}
  \arg(\det D(\mathbf{k})) .
\end{equation}
It has been shown \cite{SBT12} that this winding number can be expressed in terms of the signs of the gap functions near the Fermi surfaces as
\begin{equation}
w(\mathbf{k}_\|) = \sum_{\nu\in\lbrace +,-\rbrace}\sum_{\mathbf{k}_F^\nu} \text{sgn}(\partial_{k\perp}\,
  \xi^\nu_{\mathbf{k}}|_{\mathbf{k}=\mathbf{k}_F^\nu})\, \text{sgn}(\Delta^\nu_{\mathbf{k}_F^\nu}) ,
\end{equation}
where $\mathbf{k}_F^\nu = (\mathbf{k}_\|,k_{\perp F}^\nu)$ is the Fermi momentum of the Fermi surface $\nu$ for parallel component $\mathbf{k}_\|$.
The winding number $w(\mathbf{k}_\|)$ is therefore zero outside of the projections of the nodal lines onto the sBZ and changes by either $+1$ or $-1$ when the projection of a nodal line is crossed.
Since the flat-band surface states are protected by a topological winding number relying on chiral symmetry, which is realized as the product of particle-hole and time-reversal symmetry, we expect them to persist even in the presence of nonmagnetic impurities.

For our numerical calculations, we assume a model system with the point group $C_{4v}$ because the resulting nodal structure is rather simple.
However, any other choice of a noncentrosymmetric point group would also be valid and lead to qualitatively similar results as long as the parameters are chosen in such a way that the system has nodal lines in the bulk. While the point group determines the symmetry of the nodal lines \cite{BST11}, they exist for any noncentrosymmetric point group for appropriately chosen parameters.

\begin{figure}[tbp]
\centering
\includegraphics[width=0.35\textwidth]{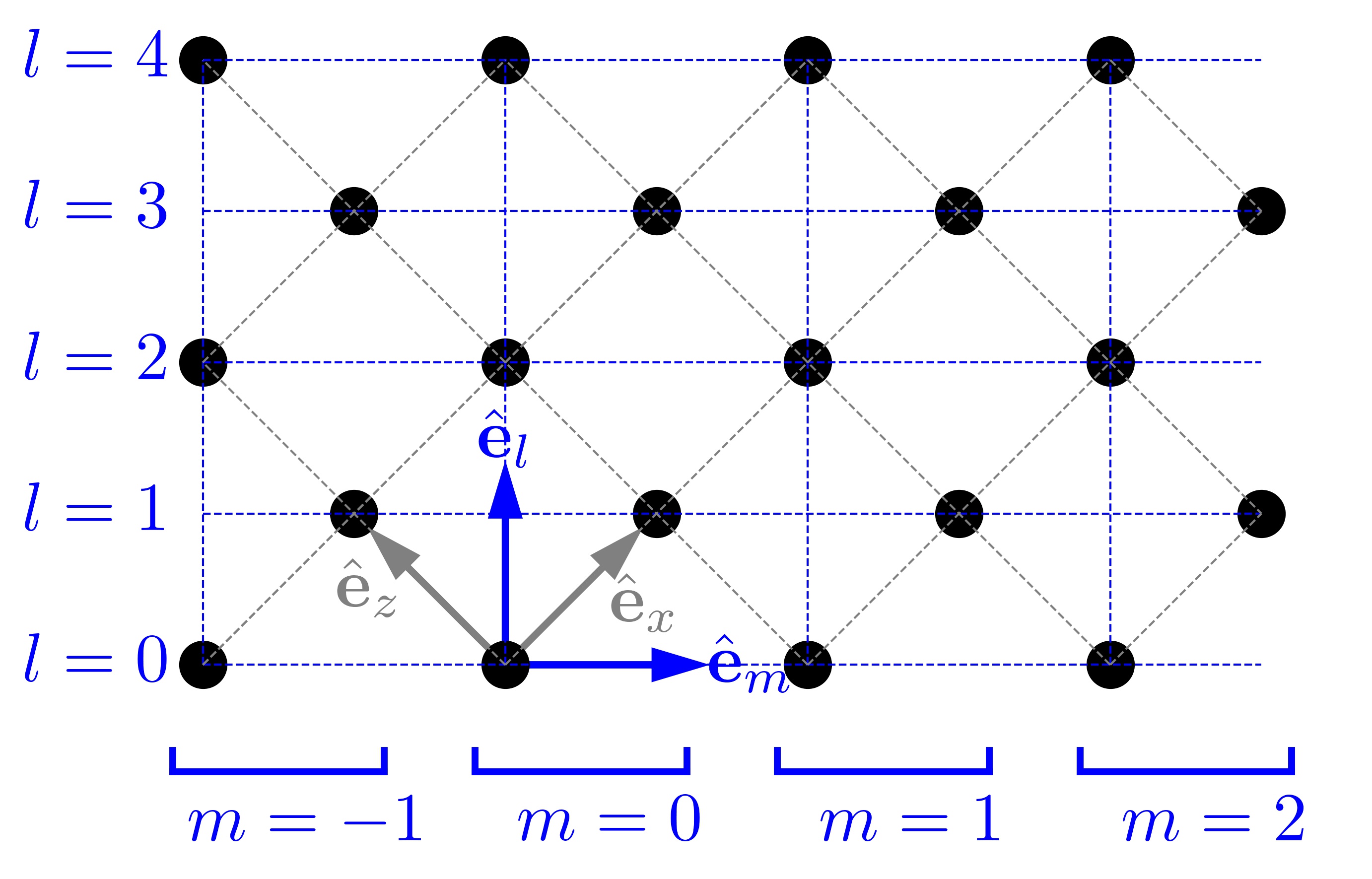}
\caption{New parallel coordinate $m$ and orthogonal coordinate $l$ at fixed $y$ for a (101) slab.}
\label{fig:coordinates}
\end{figure}

We further assume that the normal-state dispersion only contains nearest-neighbor hopping terms with amplitude $t$ such that
\begin{equation}\label{eq:dispersionepsilon}
\epsilon_{\mathbf{k}} = -2 t (\cos k_x +\cos k_y +\cos k_z) -\mu ,
\end{equation}
where $\mu$ is the chemical potential. We expand the SOC vector to lowest order, which leads to a Rashba-type SOC term
\begin{equation}\label{eq:SOC_C4v}
\mathbf{l_k} = \sin k_y\, \hat{\mathbf{e}}_x - \sin k_x\, \hat{\mathbf{e}}_y.
\end{equation}
We choose the surface orientation (101), and define the new coordinates
\begin{equation}
l = x+z
\end{equation}
orthogonal to the surface and
\begin{equation}
m = \left\lfloor \frac{x-z}{2} \right\rfloor
\end{equation}
parallel to the surface, as sketched in Fig.~\ref{fig:coordinates}. The general ideas do not depend on our particular model and calculations can be done for any point group lacking inversion and any dispersion compatible with it.

We note that for a lattice parameter $a=1$, the displacement of lattice points in the $l$ directions is $l/\sqrt{2}$ and in the $m$ direction $(2m+ l\text{ mod } 2)/\sqrt{2}$. If we define the momenta
\begin{equation}\label{eq:kl}
k_l=(k_x+k_z)/\sqrt{2}
\end{equation}
in the direction orthogonal to the surface and
\begin{equation}\label{eq:km}
k_m=(k_x-k_z)/\sqrt{2}
\end{equation}
parallel to the surface Fourier transformations along these directions therefore contain terms of the form $\exp(i k_l l/\sqrt{2})$ and $\exp[i k_m (2m+ l~\text{mod}~2)/\sqrt{2}]$, respectively.

\begin{figure}[htbp]
\centering
\includegraphics[width=0.48\textwidth]{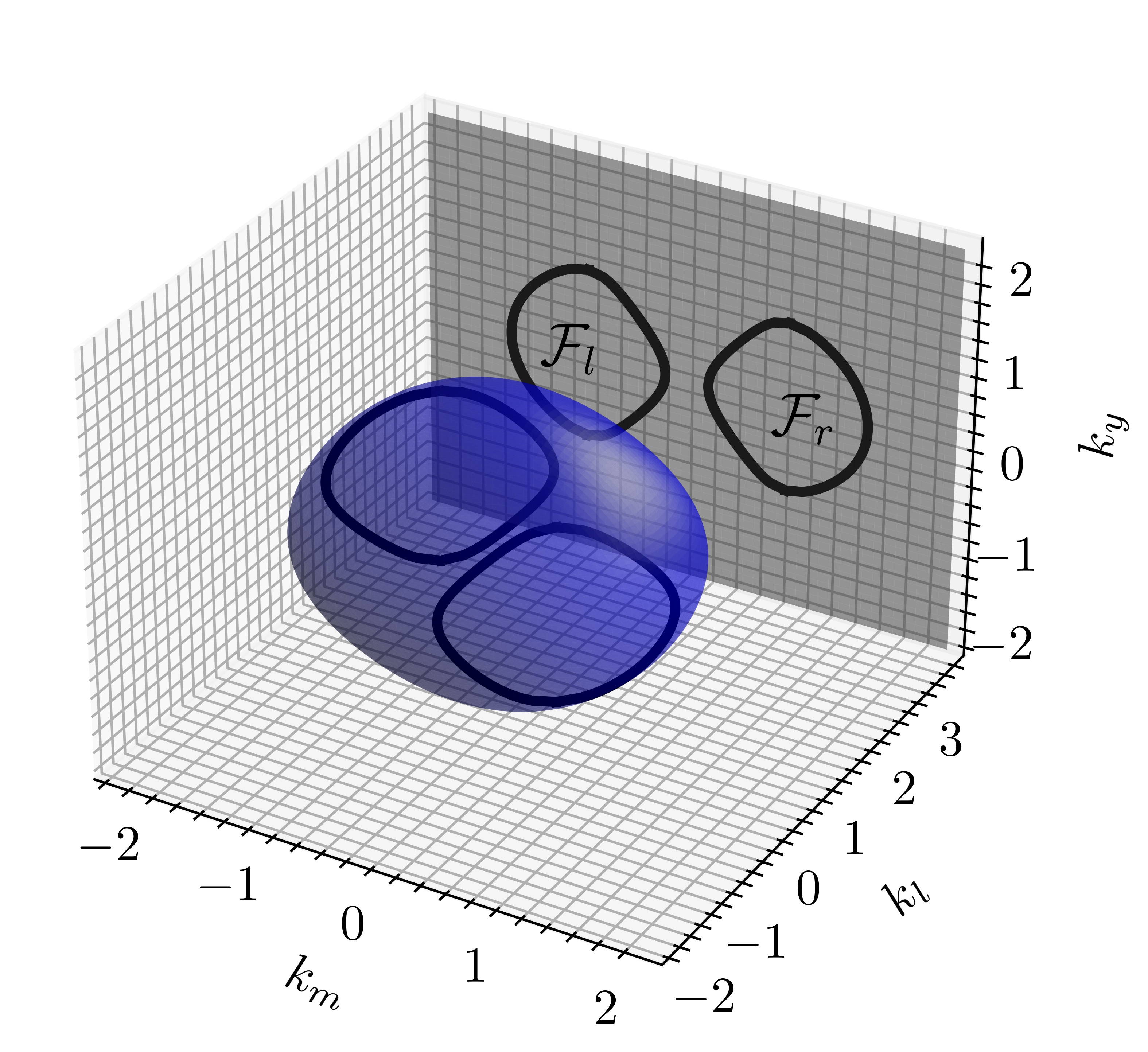}
\caption{Sketch of the Fermi surface with nodal lines and the projection onto the sBZ. The regions with nonzero winding number are labeled as $\mathcal{F}_l$ for $k_m<0$ and $\mathcal{F}_r$ for $k_m>0$.}
\label{fig:sketch_blobs}
\end{figure}

For a numerically exact diagonalization, we construct a slab Hamiltonian from Eq.~\eqref{eq:bdg_hamiltonian} with a surface at $l=0$. The derivation is presented in Appendix~\ref{sec:appendix_1}. For the numerical calculations, we use the parameters $t=1$, $\mu=-4$, $\lambda=0.05$, $\Delta^s=0.04$, and $\Delta^t=0.05$. These parameters have the dimension of energy and are given in arbitrary units. For these parameters,  the spheroidal Fermi surface has two nodal lines, which are parallel to the $xy$ plane. The projections of these lines onto the sBZ delimit two regions, which we will denote by $\mathcal{F}_\text{l}$ for the left region with $k_m<0$ and $\mathcal{F}_\text{r}$ for the right one with $k_m>0$, see Fig.~\ref{fig:sketch_blobs}.
Any other choice of parameters would also work, as long as there are nodal lines. The exact choice of parameters would of course affect the shape of the momentum region hosting zero-energy surface modes and their spin polarization.

In order to move the wave packet, we have to introduce a time-reversal-symmetry-breaking term into the Hamiltonian. We achieve this by bringing the surface into contact with a FI with a magnetization $\mathbf{h}$ that can be manipulated adiabatically. This insulator is represented by a normal-state Hamiltonian of the form
\begin{equation}\label{eq:normal_state_FI}
h_\text{FI}(\mathbf{k})=(\epsilon_{\mathbf{k}}+V) \sigma_0 +\mathbf{h}\cdot \boldsymbol{\sigma},
\end{equation}
i.e., for the sake of simplicity, we assumed the same dispersion $\epsilon_{\mathbf{k}}$ as in the superconductor and a constant potential $V$. For the numerical calculations, we will use $V=3.5$. This parameter has to be chosen large enough to ensure the presence of an energy gap so that the surface states decay exponentially into the FI. As long as this is the case, changing $V$ only results in a change of the decay length. Similarly, qualitative changes of $\epsilon_\mathbf{k}$ are also expected to only affect the decay length, as long as the gap remains open. The details of the model are described in Appendix~\ref{sec:appendix_1}.
We denote the surface state of the BdG Hamiltonian $\mathcal{H}_\text{slabs}(\mathbf{k}_\|)$ of the NCS-FI slab heterostructure with $\psi(\mathbf{k}_\|)=(\psi_{l=-L^\text{FI}}(\mathbf{k}_\|)^T,\dots,\psi_{l=L^\text{NCS}-1}(\mathbf{k}_\|)^T)^T$, i.e., $\psi(\mathbf{k}_\|)$ is a  vector of length $4(L^\text{FI}+L^\text{NCS})$ that solves the eigenvalue equation $\mathcal{H}_\text{slabs}(\mathbf{k}_\|)\, \psi(\mathbf{k}_\|)=E_{\mathbf{k}_\|} \psi(\mathbf{k}_\|)$ and $\psi_{l}(\mathbf{k}_\|)$ are vectors of length $4$.

\begin{figure}[tbp]
\centering
\includegraphics[width=0.48\textwidth]{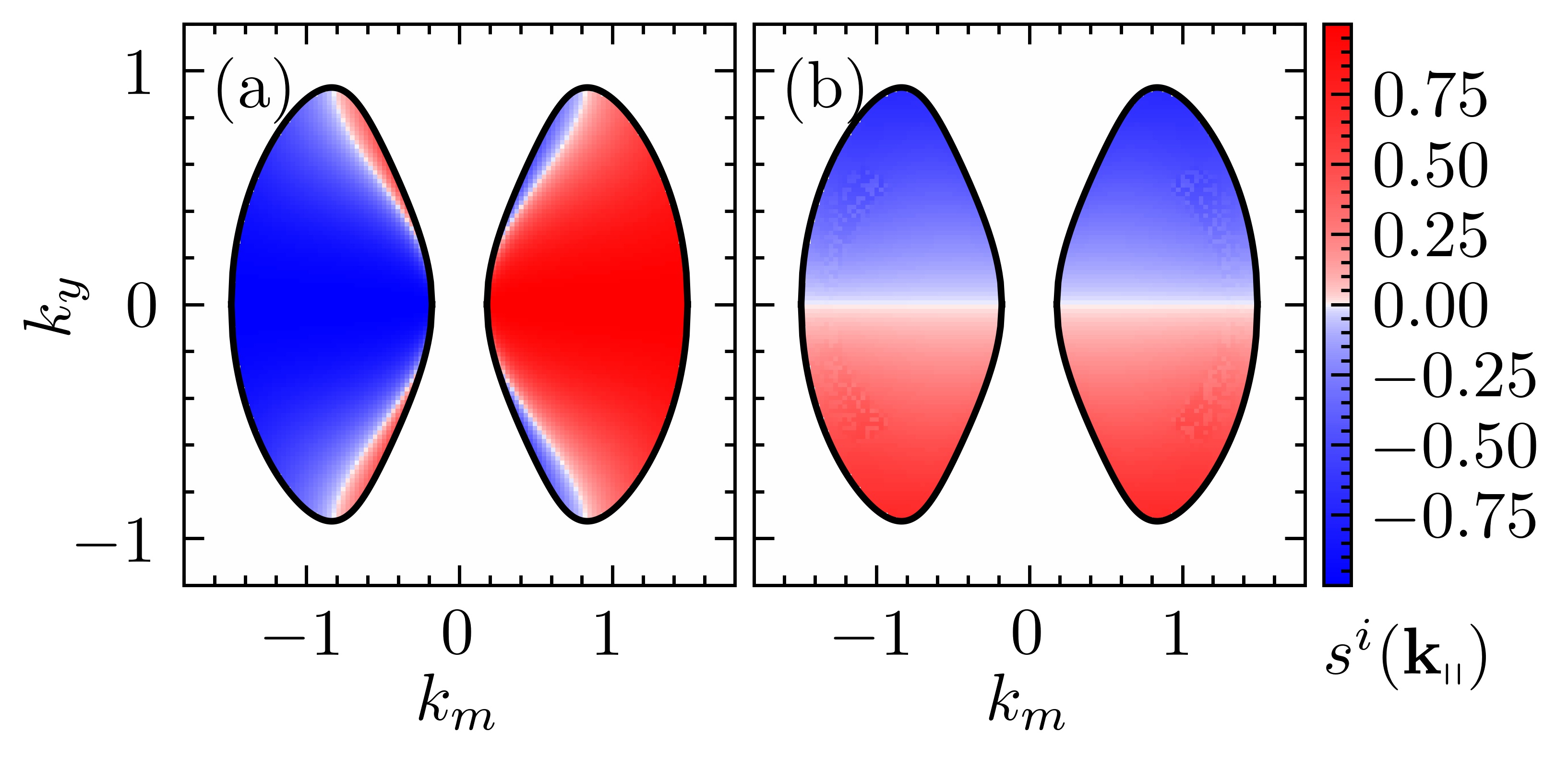}
\caption{(a) $y$ component and (b) $m$ component of the spin polarization of the surface states on the (101) surface of a NCS with the point group $C_{4v}$.}
\label{fig:spinpol}
\end{figure}

Using the spin matrices
\begin{equation}
\Sigma^i=\text{diag}(\sigma_i,-\sigma_i^T) ,
\end{equation}
we calculate the spin polarizations
\begin{equation} \label{eq:spinpol}
s^i(\mathbf{k}_\|)= \sum_l \psi_{l}(\mathbf{k}_\|)^\dagger  \Sigma^i  \psi_{l}(\mathbf{k}_\|)
\end{equation}
of the surface states $\psi(\mathbf{k}_\|)$ for the time-reversal-symmetric case of $\mathbf{h}=0$. Plots of $s^y$ and $s^m$ are shown in Fig.~\ref{fig:spinpol} for a slab thickness of $L^\text{NCS}= 5000$ for the superconducting layer and $L^\text{FI}= 200$ for the insulator. The third component $s^l$ vanishes. Since both the $y$ component $s^y(\mathbf{k}_\|)$ in Fig.~\ref{fig:spinpol}(a) and the $m$ component $s^m(\mathbf{k}_\|) = [s^x(\mathbf{k}_\|)-s^z(\mathbf{k}_\|)]/\sqrt{2}$ in Fig.~\ref{fig:spinpol}(b) are nontrivial the introduction of a magnetization $\mathbf{h}$ coupling to the spin polarization leads to a non-trivial dispersion.

\section{Construction of a wave packet}
\label{sec:Construction of a wave packet}

The goal of this section is to construct a localized wave packet as a superposition of states from the regions $\mathcal{F}_l$ and $\mathcal{F}_r$ of the sBZ and to calculate its real-space representation. As these regions only cover a fraction of the sBZ it is impossible to construct a wave packet that is localized at a single lattice site \cite{SBT12, RRT20}. The minimum size of a wave packet is inversely proportional to the momentum-space area of $\mathcal{F}_l\cup\mathcal{F}_r$.
We therefore expect that experimentally any sufficiently local excitation of a zero-energy surface-state wave packet, e.g., by tunneling from a superconducting STM tip, would result in a maximally localized wave packet. Such a wave packet consists of a linear combination of all states in $\mathcal{F}_l \cup \mathcal{F}_r$. In Sec.~\ref{sec:C4v results}, we will also consider superpositions of fewer surface states since this idea will turn out to be beneficial for the stability of the wave packet. In this section, we therefore do not assume the region $\mathcal{F} \subseteq \mathcal{F}_l \cup \mathcal{F}_r$ from which the linear combination is constructed to be the entirety of $\mathcal{F}_l \cup \mathcal{F}_r$.
The experimental implementation of a momentum-selective setup, e.g., with $\mathcal{F}\subset \mathcal{F}_l\cup\mathcal{F}_r$, will surely be demanding and is left to future research as our goal is only to give a proof of concept for the theoretical framework.

The wave packet reads as
\begin{align}\label{eq:wavepacket_sum}
&\Psi^\text{WP}(m, y, l) \notag \\
&\;=\!\! \sum_{(k_m, k_y)\in \mathcal{F}} \!\!
  e^{i \phi_{\mathbf{k}_\|}} e^{i  k_m (2m+l~\text{mod}~2)/\sqrt{2}  +i k_y y}\,
   \psi_l(\mathbf{k}_\|),
\end{align}
with phases $\phi_{\mathbf{k}_\|}$.
We choose these phases as
\begin{equation}\label{eq:phase}
\phi_{\mathbf{k}_\|} = \arg[ \psi(\mathbf{k}_\|)^\dagger v^{(0)}],
\end{equation}
where $v^{(0)}_l=(0,0,0,0)^T$ if $l\neq 0$ and $v^{(0)}_{l=0}=(1,0,1,0)^T$. This phase choice ensures that the resulting wave packet has the Majorana property, i.e., that
\begin{equation}\label{eq:Majorana_property}
\Phi_{\mathbf{R}_\|}^{\phantom{\dagger}} = \Phi_{\mathbf{R}_\|}^\dagger
\end{equation}
for the operator
\begin{equation}\label{eq:phi_operator}
\Phi_{{\mathbf{R}_\|}}=\sum_{\mathbf{k}_\|\in\mathcal{F}}e^{i \mathbf{k}_\|\cdot {\mathbf{R}_\|}} \,
  \zeta_{\mathbf{k}_\|} ,
\end{equation}
where ${\mathbf{R}_\|}=(2m + l\mod 2)/\sqrt{2}\: \hat{\mathbf{e}}_m + y\, \hat{\mathbf{e}}_y$ is the real-space position of the wave packet and
\begin{equation}
\zeta_{\mathbf{k}_\|}
  = e^{-i \phi_{\mathbf{k}_\|}}
  \sum_{l=-L^\text{FI}}^{L^\text{NCS}} \psi_{l}^\dagger(\mathbf{k}_\|) \Psi_{(\mathbf{k}_\|,l)}
\end{equation}
is the annihilation operator of the zero-energy mode at $\mathbf{k}_\|$ with the spinor
\begin{equation}
\Psi_{(\mathbf{k}_\|,l)} = (c_{(\mathbf{k}_\|,l),\uparrow},c_{(\mathbf{k}_\|,l),\downarrow},
c_{(-\mathbf{k}_\|,l),\uparrow}^\dagger,c_{(-\mathbf{k}_\|,l),\downarrow}^\dagger)^T .
\end{equation}
Note that $\Psi^\text{WP}(m, y, l)$ is the wave function corresponding to the operator $\Phi_{\mathbf{R}_\|}$.

We now show that Eq.\ (\ref{eq:Majorana_property}) holds. The PHS defined in Eq.~\eqref{eq:PHS} implies that there always exist phases $\beta_{\mathbf{k}_\|}\in\mathbb{R}$ such that
\begin{equation}\label{eq:phs}
\psi_l(\mathbf{k}_\|)=\exp(i \beta_{\mathbf{k}_\|})\, U_C\, \psi_l^\ast(-\mathbf{k}_\|).
\end{equation}
Multiplying this equation by $(v^{(0)})^\dagger=(v^{(0)})^T$ and using that $U_C\, v^{(0)}=v^{(0)}$, we find
\begin{equation}
-\phi_{\mathbf{k}_\|}=\beta_{\mathbf{k}_\|} + \phi_{-\mathbf{k}_\|}.
\end{equation}
This leads to
\begin{align}
\zeta^\dagger_{-\mathbf{k}_\|} &= e^{i \phi_{-\mathbf{k}_\|}}\sum_{l=-L^\text{FI}}^{L^\text{NCS}} \Psi^\dagger_{(-\mathbf{k}_\|,l)}\psi_l(-\mathbf{k}_\|)\notag\\
&= e^{i \phi_{-\mathbf{k}_\|}}\sum_{l=-L^\text{FI}}^{L^\text{NCS}} \Psi^T_{(\mathbf{k}_\|,l)}U_C^\dagger e^{i \beta_{\mathbf{k}_\|}}U_C\psi_l^\ast(\mathbf{k}_\|)\notag\\
&= e^{i \phi_{-\mathbf{k}_\|}}e^{i \beta_{\mathbf{k}_\|}}\sum_{l=-L^\text{FI}}^{L^\text{NCS}} \Psi^T_{(\mathbf{k}_\|,l)} \psi_l^\ast(\mathbf{k}_\|)\notag\\
&=\zeta_{\mathbf{k}_\|},
\end{align}
which, together with Eq.~\eqref{eq:phi_operator}, proves the Majorana property.

In Sec.~\ref{sec:approximation}, we develop a method to find the surface states $ \psi(\mathbf{k}_\|) $ in the limit of infinite slab thickness. In Sec.~\ref{sec:timeev}, we then examine their time evolution in real space under the influence of an adiabatically changing magnetization $\mathbf{h}$ in the FI.

\subsection{Limit of the eigenstates in momentum space for infinitely thick slabs}
\label{sec:approximation}

In this section, we consider the slab Hamiltonian for the NCS-FI heterostructure. The solution is specific to our $C_{4v}$ model system but the method can be applied to any slab heterostructure.

We note that the eigenvalue equation
\begin{equation}
\mathcal{H}_\text{slabs}(\mathbf{k}_\|)\, \psi(\mathbf{k}_\|)=E_{\mathbf{k}_\|}\psi(\mathbf{k}_\|)
\end{equation}
can be rewritten as a recurrence relation
\begin{equation}\label{eq:recurrence_NCS}
0=(H_1^\text{NCS})^\dagger\psi_{l-1} + (H_0^\text{NCS}- E \boldone_4)\psi_{l}+ H_1^\text{NCS} \psi_{l+1}
\end{equation}
on the NSC side, i.e., for $l\geq 1$, and
\begin{equation}\label{eq:recurrence_FI}
0=(H_1^\text{FI})^\dagger \psi_{l-1} + (H_0^\text{FI}- E \boldone_4)\psi_{l}+ H_1^\text{FI} \psi_{l+1}.
\end{equation}
on the FI side, i.e., for $l\leq -1$. The specific form of $H_0^\text{NCS}$, $H_1^\text{NCS}$, $H_0^\text{FI}$, and $H_1^\text{FI}$ can be found in Appendix~\ref{sec:appendix_1}. Here, we have dropped the dependence of all quantities on $\mathbf{k}_\|$ for the sake of readability.
These two equations are equivalent to the original eigenvalue equation if one imposes the boundary condition
\begin{equation}\label{eq:boundary}
0=(H_1^\text{FI})^\dagger \psi_{-1} + (H_0^\text{NCS}- E \boldone_4)\psi_{0}+ H_1^\text{NCS} \psi_{1}
\end{equation}
for the interface layer $l=0$ and the two boundary conditions
\begin{align}
  0&=(H_1^\text{NCS})^\dagger \psi_{L^\text{NCS}-2} + (H_0^\text{NCS}- E \boldone_4)\psi_{L^\text{NCS}-1} \notag \\
  &= - H_1^\text{NCS} \psi_{L^\text{NCS}}, \label{eq:top_surface}\\
  0&=(H_0^\text{FI}- E \boldone_4)\psi_{-L^\text{FI}}+ H_1^\text{FI} \psi_{-L^\text{FI}+1} \notag \\
  &=- (H_1^\text{FI})^\dagger\psi_{-L^\text{FI}-1}  \label{eq:bottom_surface}
\end{align}
for the top and bottom surface of the slab heterostructure, respectively.
However, in the following, we instead consider the limit of infinitely thick slabs, which means that Eqs.~\eqref{eq:top_surface} and~\eqref{eq:bottom_surface} are replaced by the condition that the surface state decays for $l\rightarrow \pm \infty$.
Note that in this limit, Eqs.~\eqref{eq:recurrence_NCS}--\eqref{eq:boundary} describe the system exactly.

We rewrite Eqs.~\eqref{eq:recurrence_NCS} and~\eqref{eq:recurrence_FI} employing the transfer matrices
\begin{align}
&T^\text{NCS} \notag \\
&= \begin{pmatrix}
0 & \boldone_4\\
-(H_1^\text{NCS})^{-1} (H_1^\text{NCS})^\dagger& -(H_1^\text{NCS})^{-1}(H_0^\text{NCS}- E \boldone_4)
\end{pmatrix}, \label{eq:transfermatrix_NCS}\\
&T^\text{FI}=
\begin{pmatrix}
0 & \boldone_4\\
-(H_1^\text{FI})^{-1} (H_1^\text{FI})^\dagger& -(H_1^\text{FI})^{-1}(H_0^\text{FI}- E \boldone_4)
\end{pmatrix} \label{eq:transfermatrix_FI}
\end{align}
and the vectors
\begin{equation}
x_l=\begin{pmatrix}\psi_l \\ \psi_{l+1}\end{pmatrix}
\end{equation}
as
\begin{align}
x_l &= T^\text{NCS} x_{l-1} = (T^\text{NCS})^{l}  x_0 &&\text{for }l\geq 1,\label{eq:reccurence_matrix_NCS}\\
x_l &= T^\text{FI} x_{l-1}
{} = (T^\text{FI})^{-|l|+1} x_{-1} &&\text{for }l\leq -1.\label{eq:reccurence_matrix_FI}
\end{align}
From these equations, it follows that $\psi_l$ for all $l$ can be determined from $\psi_0$ by expanding $x_0$ in the eigenbasis $\lbrace t^\text{NCS}_1,\ldots, t^\text{NCS}_8\rbrace$ ($\lbrace t^\text{FI}_1,\ldots, t^\text{FI}_8\rbrace$) of $T^\text{NCS}$ ($T^\text{FI}$) for $l>0$ ($l<0$) and multiplying with the correct power of the corresponding eigenvalues $\tau^\text{NCS}_1$, \dots, $\tau^\text{NCS}_8$ ($\tau^\text{FI}_1$, \dots, $\tau^\text{FI}_8$). Note that, due to the tautological equation $\psi_l=\psi_l$ from the upper two blocks of the transfer matrices, we do not need the full eigenvectors. Instead, it is sufficient to know the first four of the eight components of each vector, which we will denote by $t^\text{NCS}_{j,u}$ ($t^\text{FI}_{j,u}$). We find
\begin{equation}\label{eq:wavefunction_transfermatrix}
\psi_l = \frac{1}{\sqrt{n}}\sum_{j=1}^{8} \begin{cases} \alpha_j^{\text{FI}}\,
  (\tau_j^{\text{FI}})^l\, t_{j,u}^{\text{FI}} & \text{for }l<0, \\[1ex]
  \alpha_j^{\text{NCS}}\, (\tau_j^{\text{NCS}})^l\, t_{j,u}^{\text{NCS}} & \text{for }l\geq 0,
\end{cases}
\end{equation}
where $n$ is a normalization factor.
To make the wave function continuous and conform with Eq.~\eqref{eq:boundary}, the coefficients $\alpha_j^{\text{FI}}$, $\alpha_j^{\text{NCS}}$ have to satisfy
\begin{align}
0 &= \sum_{j=1}^{8} \left(\alpha_j^{\text{NCS}}  t_{j,u}^{\text{NCS}}-\alpha_j^{\text{FI}} t_{j,u}^{\text{FI}} \right), \label{eq:boundary_0}\\
0 &= \sum_{j=1}^8 \big[ \alpha_j^{\text{FI}} (\tau_j^{\text{FI}})^{-1} (H_1^\text{FI})^\dagger t_{j,u}^{\text{FI}} \notag \\
&\qquad {}+ \alpha_j^{\text{NCS}}  (H_0^\text{NCS}- E \boldone_4)\, t_{j,u}^{\text{NCS}} \notag \\
&\qquad {}+ \alpha_j^{\text{NCS}}   \tau_j^{\text{NCS}} H_1^\text{NCS} t_{j,u}^{\text{NCS}}
  \big]. \label{eq:boundary_1}
\end{align}
With Eqs.~\eqref{eq:wavefunction_transfermatrix},~\eqref{eq:boundary_0}, and~\eqref{eq:boundary_1}, the surface state for any $\mathbf{k}_\|$ can, in principle, be determined numerically. However, some further simplifications are possible, which are discussed in Appendix~\ref{sec:appendix_2}. For vanishing field, $\mathbf{h}=0$, the energy for $\mathbf{k}_\|$ inside the projections of the nodal lines is found to be $E_{\mathbf{k}_\|}=0$, as expected. This serves as a sanity check for the transfer-matrix approach. A derivation of this fact is given in Appendix~\ref{sec:appendix_3}. We also give numerical evidence that the transfer matrix approach results in a good approximation of the surface state of a slab with finite but large thickness in Appendix~\ref{sec:appendix_4}.

\subsection{Real-space wave packet and time evolution}
\label{sec:timeev}

\begin{figure}[tbp]
\centering
\includegraphics[width=0.48\textwidth]{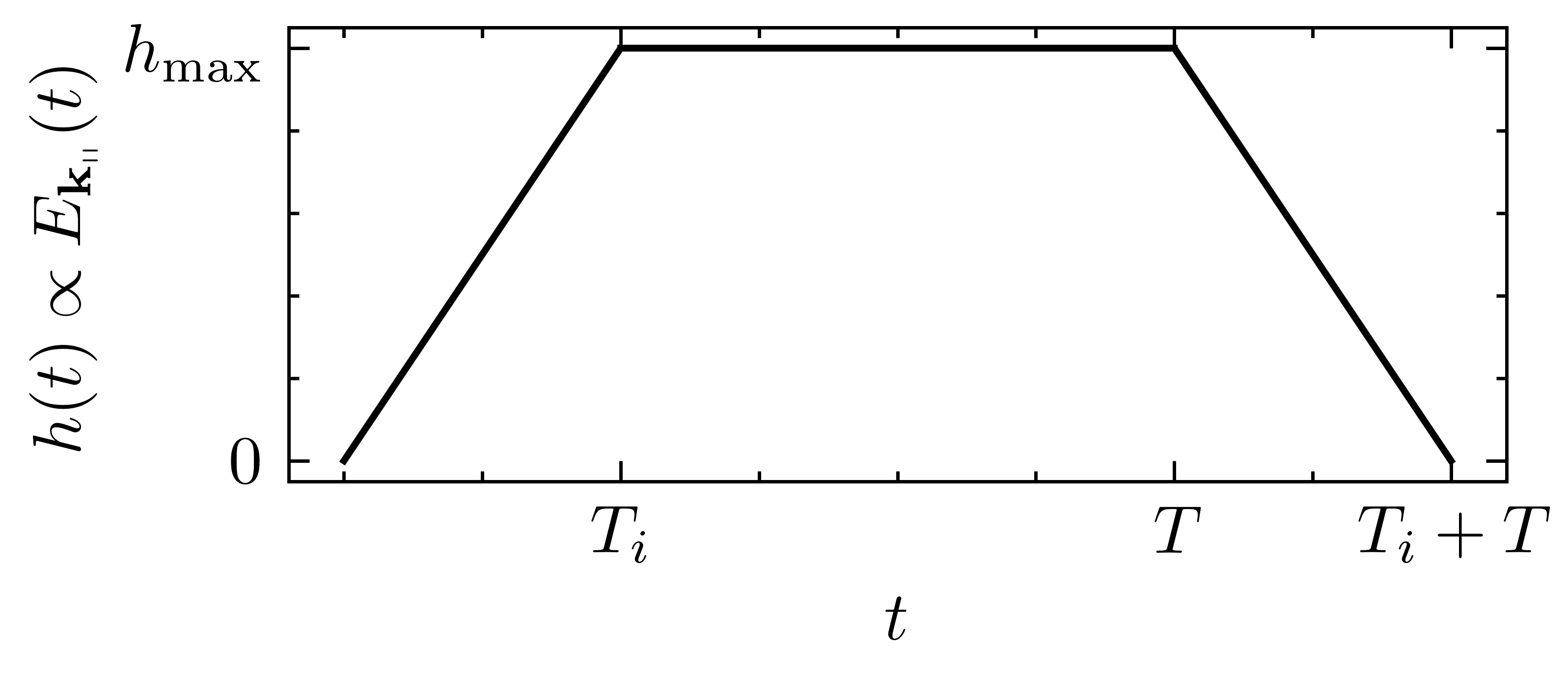}
\caption{Sketch of the time dependence of the modulus of the magnetization of the FI as given by Eq.~\eqref{eq:ht}.}
\label{fig:sketch_ht}
\end{figure}

In this section, we present the time evolution of the wave packet given in Eq.\ \eqref{eq:wavepacket_sum}. We assume that the slab is infinitely large in the parallel directions so that we can replace the sum over $\mathbf{k}_\|$ by an integral,
\begin{equation}
\sum_{(k_y, k_m)\in \mathcal{F}} \rightarrow A(\mathcal{F})\frac{\sqrt{2} }{(2\pi)^2}
  \int_{\mathcal{F}} dk_y\, dk_m ,
\end{equation}
where $A(\mathcal{F})$ is the area of $\mathcal{F}$ in momentum space. We assume that the magnetization $\mathbf{h}$ in the FI is changed adiabatically from $\mathbf{h}=0$ to $\mathbf{h}=h_\text{max}\, \hat{\mathbf{e}}$, where the direction $\hat{\mathbf{e}}$ stays constant and the change of the amplitude takes place linearly over a ramp time $T_i$. The magnetization then stays constant until $t=T$ and linearly returns to $\mathbf{h}=0$ over the time $T_i$. We thus have
\begin{equation}\label{eq:ht}
\mathbf{h}(t)=h_\text{max}\, \hat{\mathbf{e}} \times \begin{cases}
 t/T_i &\text{for }0\leq t\leq T_i,\\[0.2ex]
1 &\text{for }T_i<t< T, \\[0.2ex]
(T_i+T-t)/T_i &\text{for }T\leq t\leq T+T_i.
\end{cases}
\end{equation}
A sketch of the modulus of $\mathbf{h}(t)$ is shown in Fig.~\ref{fig:sketch_ht}.

We now examine the time evolution of the wave packet using the instantaneous eigenenergies $E_{\mathbf{k}_\|}(t)$ of the surface state at time $t$,
\begin{align}\label{eq:wavepacket_int}
&\Psi^\text{WP}(m, y, l,t) = A(\mathcal{F})\frac{\sqrt{2} }{(2\pi)^2} \notag \\
&\quad{}\times \int_{\mathcal{F}}dk_y\, dk_m\exp\left(i k_m \frac{2m+l~\text{mod}~2}{\sqrt{2}}
  + i k_y y\right) \notag \\
&\qquad{}\times \exp(i \phi_{\mathbf{k}_\|}) \exp\left(-i \int_0^t \text{d} t^\prime E_{\mathbf{k}_\|}(t^\prime)\right) \notag \\
&\qquad{}\times \exp\left(\phi^\text{geo}_{\mathbf{k}_\|}(t)\right) \psi_l(\mathbf{k}_\|,t),
\end{align}
where we have set $\hbar=1$.
In this expression, the geometric phases are given by
\begin{align}
\phi^\text{geo}_{\mathbf{k}_\|}(t)&=-\int_0^t dt^\prime\, \psi(\mathbf{k}_\|,t^\prime)^\dagger \partial_{ t^\prime} \psi(\mathbf{k}_\|,t^\prime)\notag \\
&=-\int_\mathcal{C} d\mathbf{h} \cdot \psi(\mathbf{k}_\|,\mathbf{h})^\dagger \nabla_{\mathbf{h}}\psi(\mathbf{k}_\|,\mathbf{h}) ,
\end{align}
where $\psi(\mathbf{k}_\|,t^\prime)$ are the instantaneous eigenstates of $\mathcal{H}(\mathbf{k}_\|,t^\prime)$ at energies $E_{\mathbf{k}_\|}(t^\prime)$ and $\mathcal{C}$ is the contour traced out by $\mathbf{h}(t)$.
The use of instantaneous eigenstates is justified if the time evolution is adiabatic. The condition of adiabaticity is discussed in more detail in Appendix \ref{sec:appendix_5}.

For every point $(m, y, l)$, the integral in Eq.~\eqref{eq:wavepacket_int} can, in principle, be evaluated numerically, using the surface states $\psi(\mathbf{k}_\|,t^\prime)$ and energies $E_{\mathbf{k}_\|}(t)$ calculated as described in Sec.~\ref{sec:approximation}. However, we instead make a few simplifying assumptions and approximations, which will lead to numerically less expensive calculations.

We begin by observing that for sufficiently small $\mathbf{h}$, the surface-state energy depends linearly on the modulus $|\mathbf{h}|$ \cite{LT22}, which is plausible due to the bilinear coupling between the magnetization and the spin polarization. This gives a linear energy dependence at the first order of perturbation theory in $\mathbf{h}$. The dynamical phase can therefore be simplified to
\begin{align}
&-\int_0^t dt\, E_{\mathbf{k}_\|}(t) = - E_{\mathbf{k}_\|}^{h_\text{max}} \notag \\
&\quad{} \times \begin{cases}
 t^2/(2T_i) &\text{for }0\leq t\leq T_i, \\[0.2ex]
 t-T_i/2 &\text{for }T_i<t< T, \\[0.2ex]
 \displaystyle -\frac{(t-T_i)^2-2 T t+T^2}{2T_i}&\text{for }T\leq t\leq T+T_i,
\end{cases}
\end{align}
where $E_{\mathbf{k}_\|}^{h_\text{max}}$ is the surface-state energy for $|\mathbf{h}|=h_\text{max}$.
To further simplify the results, we restrict our examinations to the time $t=T_i+T$, i.e., we only consider the final wave packet after the magnetization has been turned on and off again. At this time, we find
\begin{align}
-\int_0^t d t~ E_{\mathbf{k}_\|}(t) = -E_{\mathbf{k}_\|}^{h_\text{max}}(T+T_i)=-E_{\mathbf{k}_\|}^{h_\text{max}} T
\end{align}
and
\begin{align}
\phi^\text{geo}_{\mathbf{k}_\|}(T+T_i)=&
-\int_0^{T_i} dt^\prime\, \psi(\mathbf{k}_\|,t^\prime)^\dagger \partial_{ t^\prime} \psi(\mathbf{k}_\|,t^\prime) \notag \\
&-\int_{T_i}^{T} dt^\prime\, \psi(\mathbf{k}_\|,t^\prime)^\dagger \underbrace{\partial_{ t^\prime}\psi(\mathbf{k}_\|,t^\prime)}_{0} \notag \\
&-\int_T^{T+T_i} dt^\prime\, \psi(\mathbf{k}_\|,t^\prime)^\dagger \partial_{ t^\prime} \psi(\mathbf{k}_\|,t^\prime).
\end{align}
If we now substitute $t^\prime \rightarrow \tilde{t}=T+T_i- t^\prime$ in the third term we find that the geometric phase vanishes,
\begin{equation}
\phi^\text{geo}_{\mathbf{k}_\|}(T+T_i)=0.
\end{equation}
To calculate the integral in Eq.~\eqref{eq:wavepacket_int}, we thus only need the surface-state wave functions $\psi(\mathbf{k}_\|)$ at $\mathbf{h}=0$ and the energies $E_{\mathbf{k}_\|}^{h_\text{max}}$ at the maximum value of $|\mathbf{h}|$. In particular, the geometric phase, and with it any terms containing wave functions $\psi(\mathbf{k}_\|,\mathbf{h})$ and their derivatives, vanish because the magnetization is switched on and off in the same direction instead of being rotated. The dynamical phase can be written in terms of only $E_{\mathbf{k}_\|}^{h_\text{max}}$ and not $E_{\mathbf{k}_\|}(h(t))$ because we assumed the surface-state energy to depend linearly on $h$.

\section{Results for point group \texorpdfstring{$C_{4v}$}{C4v}}
\label{sec:C4v results}

In this section, we present the time evolution for a slab with the point group $C_{4v}$.
As before, the system parameters are chosen as $t=1$, $\mu=-4$, $\lambda=0.05$, $V=3.5$, $\Delta^s=0.04$, $\Delta^t=0.05$, and  $h_\text{max}=0.05$.
The numerical integration of the wave function in Eq.~\eqref{eq:wavepacket_int} is performed with a global adaptive method in Mathematica \cite{Mathematica}, setting both the precision goal and the accuracy goal to 4. We begin by examining a wave packet constructed from zero-energy states from the full region $\mathcal{F}_l\cup \mathcal{F}_r$. We obtain
\begin{align}\label{eq:wavepacket_int_full}
&\Psi^\text{WP}(m, y, l, T+T_i) = A(\mathcal{F}_r)\, \frac{\sqrt{2}}{(2\pi)^2} \notag \\
&\quad{} \times \int_{-k_y^\text{max}}^{k_y^\text{max}} dk_y
  \left(
  \int_{-k_m^\text{max}(k_y)}^{-k_m^\text{min}(k_y)} dk_m
  + \int_{k_m^\text{min}(k_y)}^{k_m^\text{max}(k_y)} dk_m
  \right) \notag \\
&\qquad{} \times \exp\left(i  k_m \frac{2m+l~\text{mod}~2}{\sqrt{2}}  +i k_y y\right) \notag \\
&\qquad{} \times \exp(i \phi_{\mathbf{k}_\|}) \exp\left(-i E_{\mathbf{k}_\|}^{h_\text{max}} T\right) \psi_l(\mathbf{k}_\|,\mathbf{h}=0).
\end{align}
The boundaries $k_m^\text{min}(k_y)$, $k_m^\text{max}(k_y)$, and $k_y^\text{max}$ of $\mathcal{F}$ can be expressed analytically for our model and are presented in Appendix~\ref{sec:appendix_6}.

\begin{figure*}[tbp]
\centering
\includegraphics[width=\textwidth]{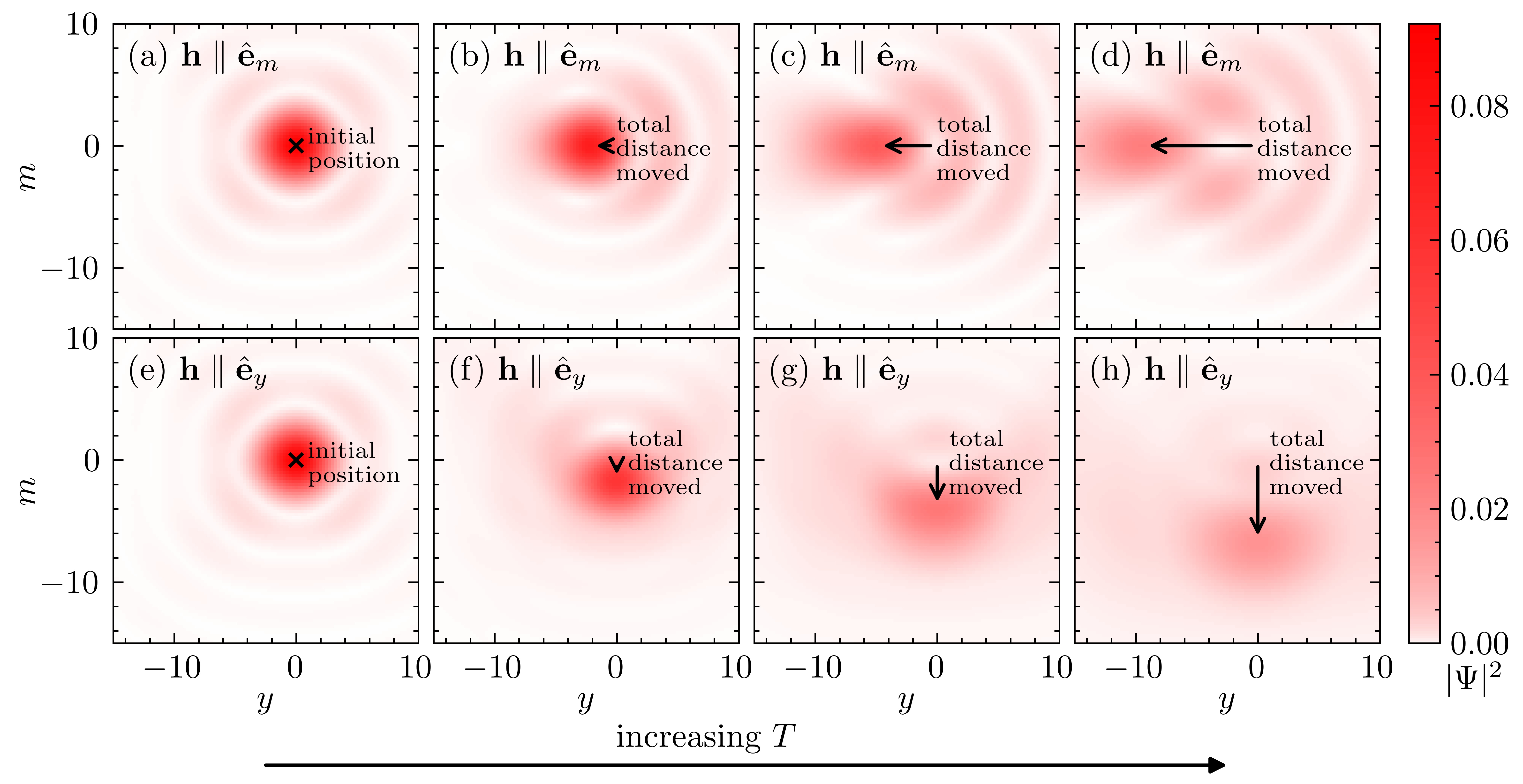}
\caption{Time evolution of a maximally localized wave packet in the $l=0$ plane. The first row shows the time-evolved packet for a magnetization in the $m$ direction at times (a) $T=0$, (b) $T=4\times 10^4$, (c) $T=8\times 10^4$, and (d) $T=1.2\times 10^5$. The second row shows the time-evolved packet for a magnetization in the $y$ direction at times (e) $T=0$, (f) $T=2\times 10^4$, (g) $T=4\times 10^4$, and (h) $T=6\times 10^4$.}
\label{fig:wavefunction_timevolved_fullarea}
\end{figure*}

Figure~\ref{fig:wavefunction_timevolved_fullarea} shows initial wave packets and several time-evolved wave packets in the layer $l=0$ for different times $T$. In the first row, panels (a)--(d), the magnetization points in the $m$ direction and the times $T$ are chosen as $T=0$, $4\times 10^4$, $8\times 10^4$, and $1.2\times 10^5$, respectively.
In the second row, panels (e)--(h), the magnetization points in the $y$ direction and we choose $T=0$, $2\times 10^4$, $4\times 10^4$, and $6\times 10^4$, respectively.
Note that the units of $T$ are the inverse of the units of energy.

It becomes clear that the wave packet is indeed spatially localized and moves in the negative $y$ direction for $\mathbf{h}\parallel \hat{\mathbf{e}}_m$ and in the negative $m$ direction for $\mathbf{h}\parallel \hat{\mathbf{e}}_y$.
Interestingly, the fact that the support $\mathcal{F}$ of the superposition in Eq.\ \eqref{eq:wavepacket_sum} is symmetric under time reversal and mirror reflections does not prevent directed motion of the wave packet.
We also see that the time evolution leads to a significant broadening of the wave packet, which increases with time $T$.
Both the direction of motion and the broadening are expected features as the exchange field couples to the spin polarization shown in Fig.~\ref{fig:spinpol}. The energies $E_{\mathbf{k}_\|}^{h_\text{max}}$ of the surface states with a magnetization in a certain direction are therefore proportional to the spin polarization in that direction. These energies determine the time evolution. A linear dispersion would lead to a motion without broadening, while nonlinearities increase the width and change the shape of the wave packet. Therefore, the wave packets broaden rather quickly as the spin polarization is strongly nonlinear in $\mathbf{k}_\|$.

\begin{figure}[tbp]
\centering
\includegraphics[width=0.48\textwidth]{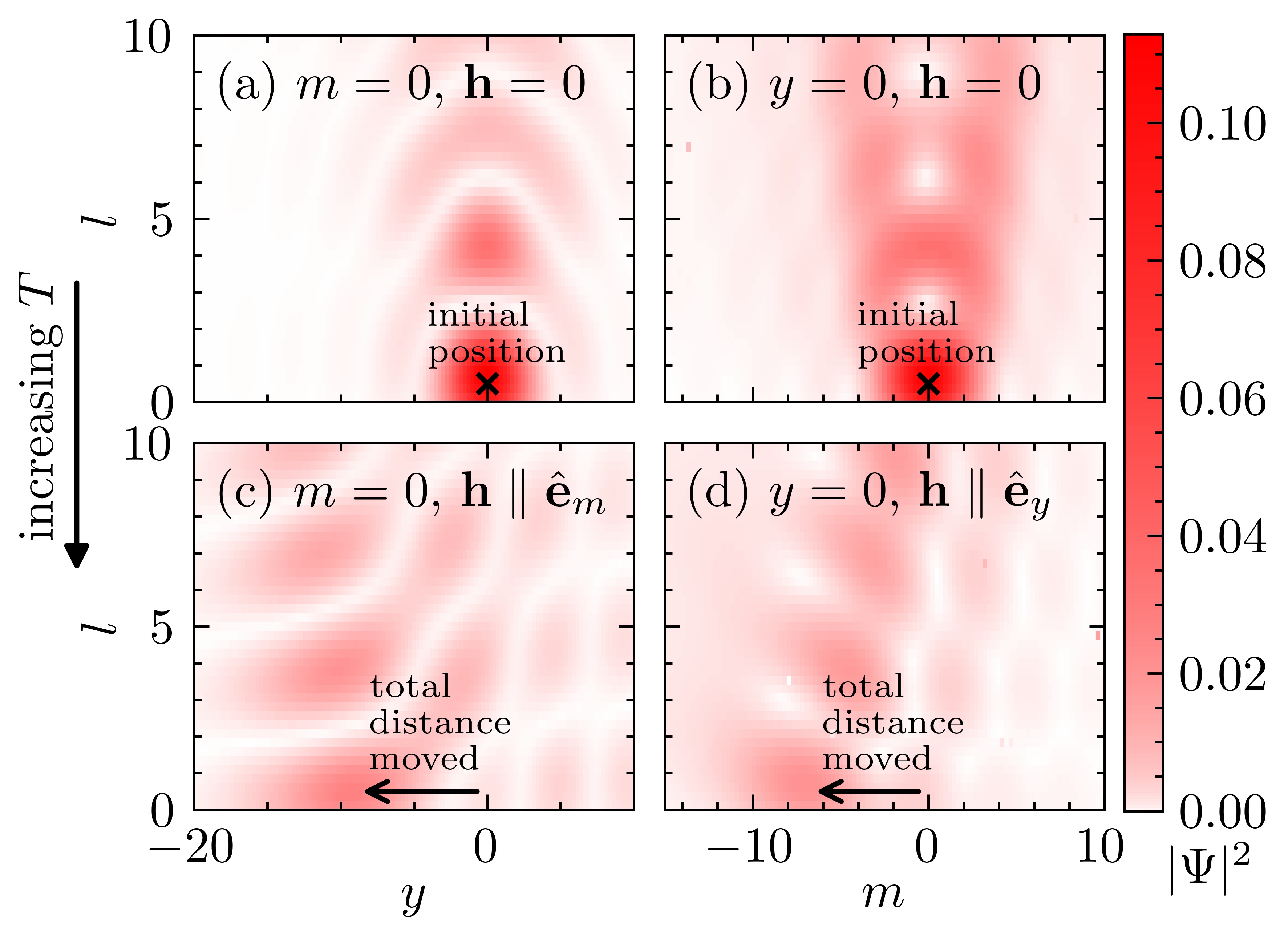}
\caption{Cuts through the NCS slab of (a), (b) the initial and (c), (d) the time-evolved wave packets. Panels (a) and (c) show cuts in the $m=0$ plane, whereas panels (b) and (d) show cuts in the $y=0$ plane. While the initial packets correspond to $T=0$, the time-evolved packets are shown for the times $T=1.2\times 10^5$ in panel (c) and $T=6\times 10^4$ in panel~(d).}
\label{fig:wavefunction_timevolved_fullarea_l}
\end{figure}

Figure~\ref{fig:wavefunction_timevolved_fullarea_l} shows cuts through the NCS slab of the wave packets at $m=0$ in Figs.~\ref{fig:wavefunction_timevolved_fullarea_l}(a) and \ref{fig:wavefunction_timevolved_fullarea_l}(c) and at $y=0$ in Figs.~\ref{fig:wavefunction_timevolved_fullarea_l}(b) and \ref{fig:wavefunction_timevolved_fullarea_l}(d). The first row, panels (a) and (b), shows the initial wave packet, while the second row, panels (c) and (d), shows time-evolved wave packets at $T=1.2\times 10^5$ and $6\times 10^4$, respectively. Both the initial and the time-evolved wave packets are localized at the surface. It is indeed expected that the wave packets remain localized under the adiabatic time evolution since the ramp time $T_i$ is long compared to the inverse bulk gap. We again observe the motion of the wave packet in the direction orthogonal to the magnetization as well as the broadening.

\begin{figure}[tbp]
\centering
\includegraphics[width=0.48\textwidth]{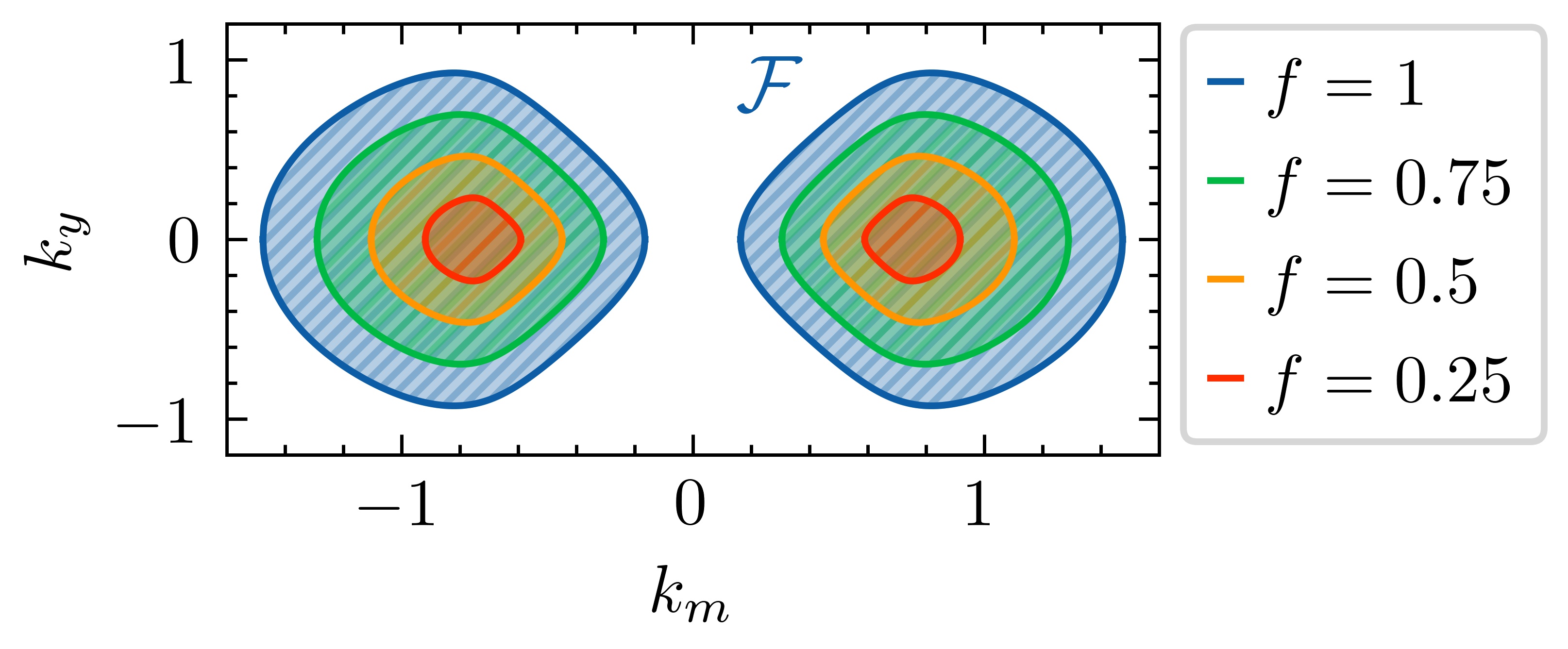}
\caption{Sketch of the smaller regions of integration $\mathcal{F}_f\subseteq \mathcal{F}_l\cup\mathcal{F}_r $}
\label{fig:blobs}
\end{figure}

How can we reduce the undesirable broadening? As mentioned in the beginning of Sec.~\ref{sec:Construction of a wave packet}, this can be achieved by restricting the superposition to smaller regions $\mathcal{F}\subseteq \mathcal{F}_l\cup \mathcal{F}_r$.
It is clear, though, that the practical generation of such a superposition would be challenging. Shrinking the support in momentum space necessarily makes the initial wave packet broader in real space, which seems to conflict with our goals. However, we will show that the increase of the width accumulated during time evolution can be reduced dramatically by this alteration.
We choose a simple and systematic way of defining the smaller region, which we will denote by $\mathcal{F}_f$. Here, $f\in(0,1]$ is a number that denotes the square root of the fraction of the area of $\mathcal{F}_l\cup \mathcal{F}_r$ over which we will integrate to construct the wave packet. The new range of integration $\mathcal{F}=\mathcal{F}_f$ again consists of two unconnected regions, which are subsets of $\mathcal{F}_l$ and $\mathcal{F}_r$. The two regions are mirror images of each other, which is necessary to preserve the Majorana property. The right-hand region is defined in such a way that its center of mass coincides with the center of mass of $\mathcal{F}_r$ and the distance from this center is scaled by a factor of $f \le 1$, i.e.,
\begin{align}
k_{y,f}^\text{max}&=f k_y^\text{max} , \\
k_{m,f}^\nu(k_y)&= k_m^\text{cent}+f\left[k_m^\nu(k_y/f)-k_m^\text{cent}\right].
\end{align}
Here, the coordinates of the center of mass are given by $k_y=0$ and
\begin{align}
k_m^\text{cent} &= \frac{\iint_{\mathcal{F}_r} k_m}{\iint_{\mathcal{F}_r}} \notag \\
&= \frac{\int_{-k_y^\text{max}}^{k_y^\text{max}} dk_y
  \int_{k_m^{\nu=1}(k_y)}^{k_m^{\nu=-1}(k_y)} dk_m\, k_m}
  {\int_{-k_y^\text{max}}^{k_y^\text{max}} dk_y
  \int_{k_m^{\nu=1}(k_y)}^{k_m^{\nu=-1}(k_y)} dk_m} \notag \\
&= \frac{1}{2}\frac{\int_{-k_y^\text{max}}^{k_y^\text{max}} dk_y \left\lbrace
  \left[k_m^{\nu=-1}(k_y)\right]^2-\left[k_m^{\nu=1}(k_y)\right]^2\right\rbrace}
  {\int_{-k_y^\text{max}}^{k_y^\text{max}} dk_y \left[k_m^{\nu=-1}(k_y)-k_m^{\nu=1}(k_y)\right]}.
\end{align}
A sketch of this is shown in Fig.~\ref{fig:blobs}.

\begin{figure*}[tbp]
\centering
\includegraphics[width=\textwidth]{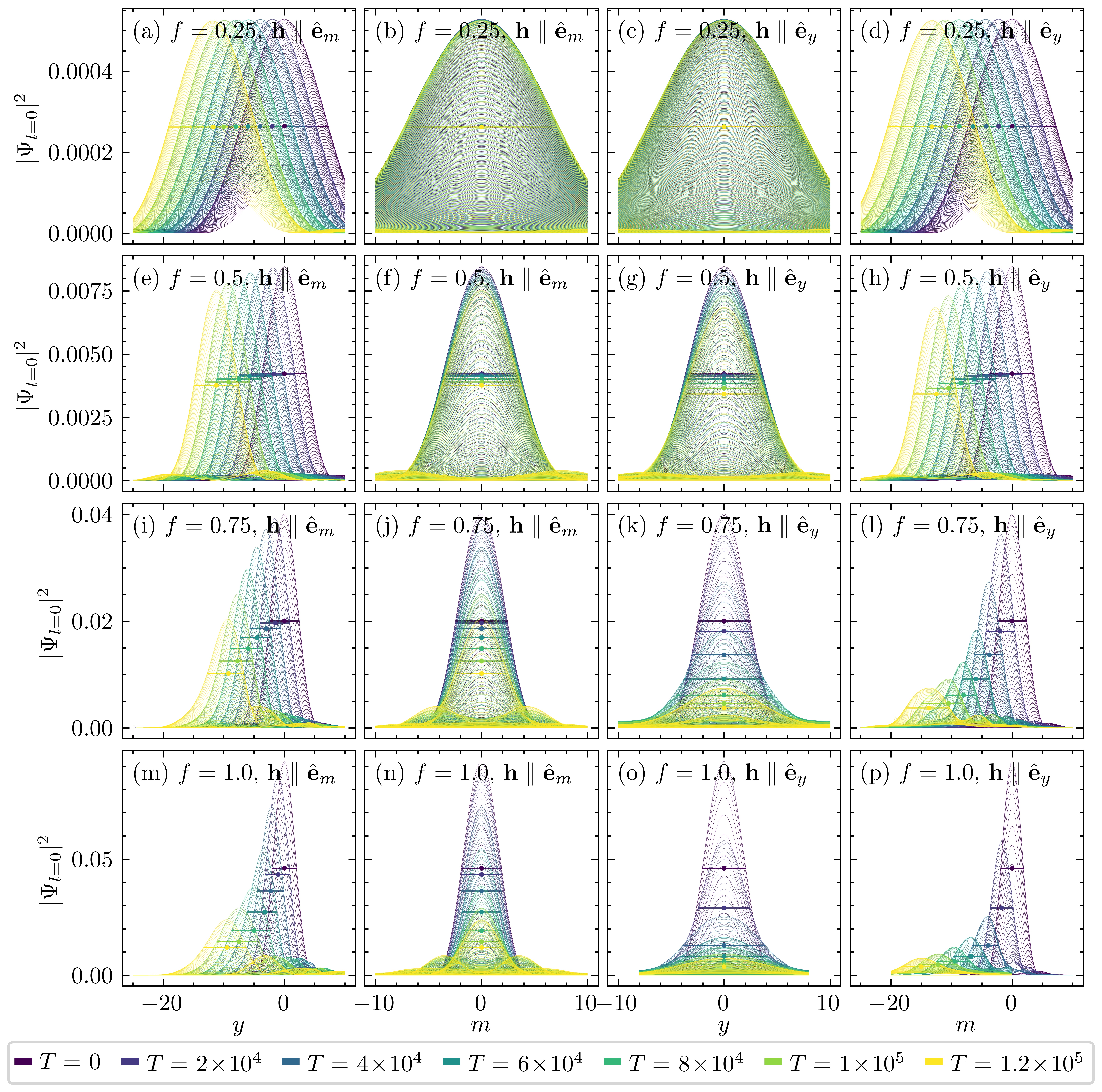}
\caption{Time evolution of wave packets which result from integrating over only a fraction of the region $\mathcal{F}_f\subseteq \mathcal{F}_l\cup \mathcal{F}_r $, shown for the top layer $l=0$. The magnetization points in the $m$ direction for the first two columns, panels (a), (b), (e), (f), (i), (j), (m), and (n) and in the $y$ direction for the last two columns, panels (c), (d), (g), (h), (k), (l), (o), and (p). The first and third column show a projection along the $m$ axis and the second and fourth column show a projection along the $y$ axis. The four rows correspond to different values of $f=0.25$, $0.5$, $0.75$, and $1$, respectively. The color scale is used to indicate the variable $T$, which is related to the time that has elapsed during the time evolution of the particular wave packet, see Fig.~\ref{fig:sketch_ht}.
}
\label{fig:wavefunction_partialarea}
\end{figure*}

Figure \ref{fig:wavefunction_partialarea} shows projections of wave packets in the layer $l=0$ with a magnetization along the $m$ direction for wave packets constructed from regions $\mathcal{F}_f$ with various scaling factors $f$. Since the motion in deeper layers $l>0$ roughly follows the one in the top layer $l=0$, as shown in Fig.~\ref{fig:wavefunction_timevolved_fullarea_l}, we here focus on the top layer. The magnetization points in the $m$ direction for the first two columns of Fig.\ \ref{fig:wavefunction_partialarea}
and in the $y$ direction for the third and fourth column.
The first and third column show projections of the wave packets along the $m$ direction and the second and fourth column show projections along the $y$ direction. The rows correspond to different sizes of $\mathcal{F}_f$, $f=0.25$, $0.5$, $0.75$, and $1$. A color scale is used to indicate the time variable $T$.
We also plot a dot at the coordinates $y^\text{cent}$ and $m^\text{cent}$ where the modulus of a wave function $|\Psi_{l=0}|^2$ reaches its maximum. The horizontal crossing the dot indicates the full width at half-maximum (FWHM) along the given direction.

We observe that for all values of $f$, the wave packets move only in the direction orthogonal to the magnetization but broaden in both directions. While the motion of $y^\text{cent}$ and $m^\text{cent}$ does not depend significantly on $f$, we see that the increase of $w(T)$ is much smaller for small values of $f$, implying that a smaller region $\mathcal{F}_f$ indeed reduces the delocalization due to the time evolution. We also see that for smaller regions $\mathcal{F}_f$, the initial wave packet is more spread out.

\begin{figure}[tbp]
\centering
\includegraphics[width=0.48\textwidth]{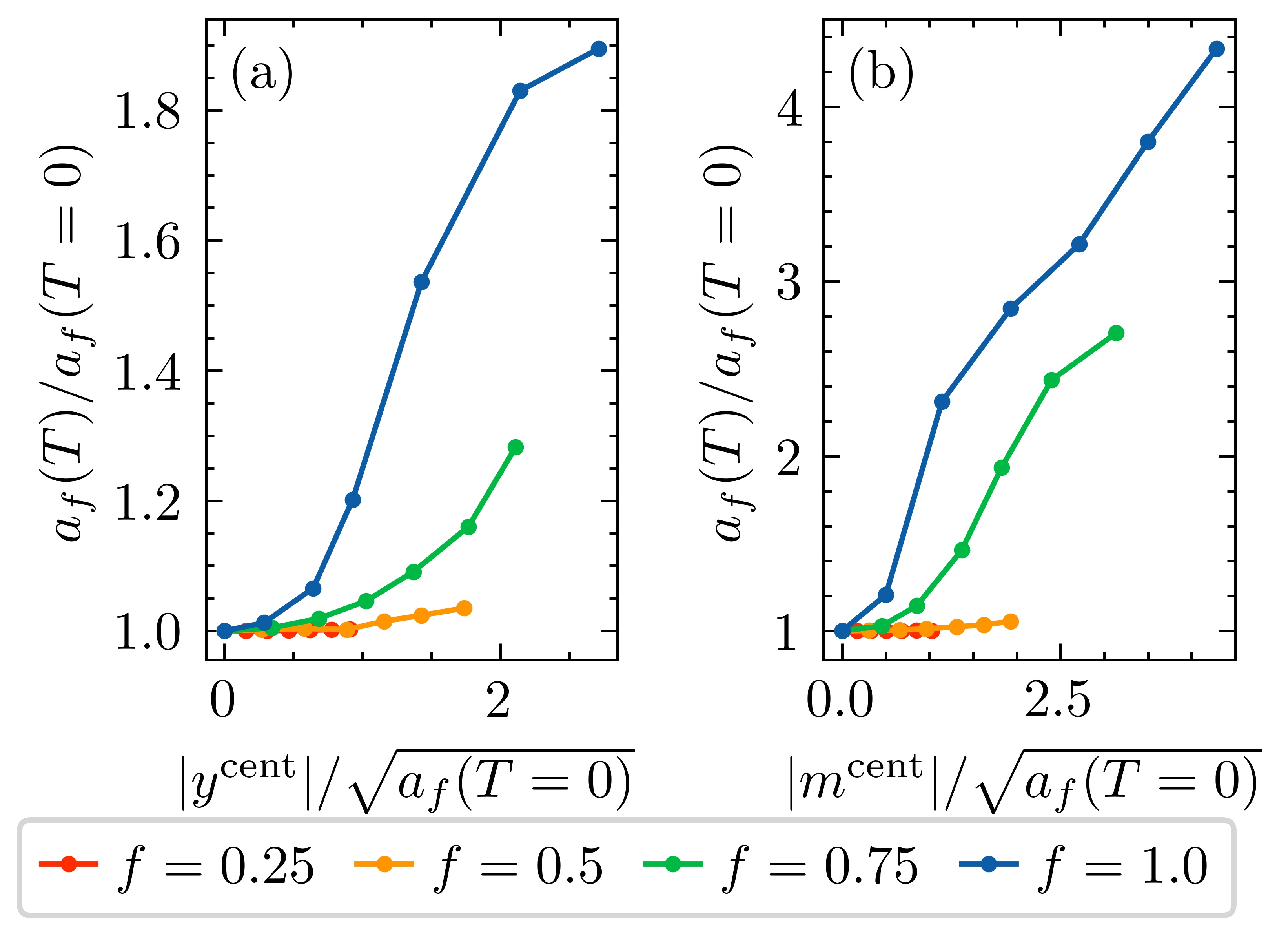}
\caption{Comparison of the motion and broadening of wave packets containing flat-band states from only a fraction of the region $\mathcal{F}_f\subseteq \mathcal{F}_l\cup \mathcal{F}_r $, in the layer $l=0$. The wave packets are time evolved with a magnetization (a) in the $m$ direction and (b) in the $y$ direction. The horizontal axis shows the absolute value of the position of the wave-packet center along the direction of motion, normalized to the square root of the initial area of the wave packet, and the vertical axis shows the area normalized to its initial value.}
\label{fig:wavefunction_comparison}
\end{figure}

Figure~\ref{fig:wavefunction_comparison} summarizes our findings concerning the motion of the wave packets for a magnetization in the $m$ direction in panel (a) and for a magnetization in the $y$ direction in panel (b). The purpose of this figure is to compare the distance moved to the width of the final wave packet for different values of $f$.
The vertical axis shows the area $a_f(T)$ in real space in which $|\Psi_{l=0}|^2$ is larger than half its maximum value. This quantity is a two-dimensional generalization of the FWHM and encompasses the broadening in both surface directions. To make this quantity comparable for different values of $f$, we normalize it to its initial value $a_f(T=0)$. The horizontal axis gives the absolute value of the coordinate of the wave-packet center in the direction of motion, normalized by the square root of $a_f(T=0)$. This means that a shallower slope of the curves corresponds to weaker delocalization. We see that for smaller values of $f$, the wave packets delocalize very little, which is evidenced by the almost horizontal red and yellow lines.

\section{Summary and conclusions}
\label{sec:summary}

In summary, we have investigated the time evolution of localized wave packets constructed from Majorana zero-energy states at the surface of NCSs. Due to the presence of zero-energy surface modes in a part of the sBZ, it is mathematically straightforward to construct a superposition of surface states that is localized in real space. Any envisaged application of such Majorana wave packets for quantum computation requires that they can be moved along the surface without excessive loss of their localization and coherence and that they exhibit non-abelian braiding statistics. In this paper, we have addressed the first point by exploring the motion of the wave packets by means of a time-reversal-symmetry-breaking exchange field. This field is introduced by bringing the NCS into contact with a FI with magnetization $\mathbf{h}$.
This is only a first step towards realizing a braiding protocol that could, e.g., exchange wave packets localized at two points in space. Beyond effects arising from the individual broadening and the overlap of a pair of wave packets, there might be a rotation in the degenerate subspace of Fock space corresponding to the zero-energy surface states when moving two wave packets around each other. However, a spatially homogeneous FI as considered in this work would only move both wave packets in the same direction. For braiding, it will instead be necessary to introduce an inhomogeneous magnetization.

We have introduced a general NCS model with a FI slab on top, as well as a specific model with the point group $C_{4v}$ and a $(101)$ surface, which we have used for our numerical calculations. After constructing a localized wave packet from the zero-energy surface states, we have established a transfer-matrix method to calculate the zero-energy surface states in the limit of infinitely thick slabs. This method is much faster and less memory intensive than numerical exact diagonalization of the BdG Hamiltonian of a thick slab but just as precise for sufficiently large slab thickness. For vanishing magnetization of the FI, $\mathbf{h}=0$, the surface states can be obtained analytically. The transfer-matrix method also allows us to calculate the surface states and the surface-state energies at $\mathbf{h}\neq 0$. While this is less efficient than for $\mathbf{h}=0$ as it requires a root-finding algorithm, it also uses less memory and is still efficient for large slab thickness.

Applying the transfer-matrix method, we have calculated the time-evolved wave packet and showed that an NCS-FI heterostructure with a magnetization in the $m$ direction moves the wave packet in the $y$ direction and vice versa. If we construct the Majorana wave packets by superposition of states from the entire sBZ region containing the flat surface band, then the states broaden rather quickly since the energy dispersion becomes strongly nonlinear. We have therefore explored an alternative method of constructing the wave packets from only part of the support of the zero-energy states, which initially leads to broader wave packets. However, these wave packets delocalize much more slowly during time evolution.

The analysis in the paper is intended as a proof of concept. We have presented a method of moving wave packets of zero-energy surface states along the surface using the magnetization of a FI. We have also shown that this motion is associated with weaker delocalization if the initial wave packet is constructed from a smaller region of the sBZ. The protocol presented here only moves a single wave packet in a certain direction. In order to move two wave packets around each other, e.g., for braiding, it would be necessary to move two wave packets in opposite directions. This cannot be achieved with the methods presented in this work because we rely on the assumption of momentum conservation and therefore assume transitional symmetry. However, this work provides a first step for studying the behavior of the Majorana wave packets during braiding.

\begin{acknowledgments}

Financial support by Deutsche Forschungsgemeinschaft, in part through Collaborative Research Center SFB 1143, project A04, project ID 247310070, and W\"urzburg-Dresden Cluster of Excellence ct.qmat, EXC 2147, project ID 390858490, is gratefully acknowledged.

\end{acknowledgments}

\appendix

\section{Slab Hamiltonian}
\label{sec:appendix_1}

In this appendix, we derive the Hamiltonian matrices for a superconducting slab with and without a magnetic insulator at its surface. We begin with the $4\times 4$ BdG Hamiltonian given in Eq.~\eqref{eq:bdg_hamiltonian} with blocks described by Eqs.~\eqref{eq:normal_state_hamiltonian}, \eqref{eq:gap_matrix}, \eqref{eq:dispersionepsilon}, and \eqref{eq:SOC_C4v}. We then use the definitions of the perpendicular momentum $k_l$ in Eq.~\eqref{eq:kl} and the parallel momentum $k_m$ in Eq.~\eqref{eq:km} and $k_y$ to transform all equations from $k_x$, $k_y$, and $k_z$ to $k_m$, $k_y$, and $k_l$, which amounts to the rotation
\begin{equation}
\begin{pmatrix}
k_x\\k_y\\k_z
\end{pmatrix}
= \begin{pmatrix}
\frac{1}{\sqrt{2}}  & 0 & \frac{1}{\sqrt{2}}\\
0                   & 1 & 0                 \\
-\frac{1}{\sqrt{2}} & 0 & \frac{1}{\sqrt{2}}
\end{pmatrix}
\begin{pmatrix}
k_m\\k_y\\k_l
\end{pmatrix}.
\end{equation}
In the next step, we Fourier transform along the perpendicular ($l$) direction,
\begin{equation}
c_{(\mathbf{k}_\|,k_l),\sigma}=\frac{1}{\sqrt{L^\text{NCS}}}\sum_{l=0}^{L^\text{NCS}-1}\exp\left({\frac{-i k_l l}{\sqrt{2}}}\right)c_{(k_m,k_y,l),\sigma} ,
\end{equation}
where we introduce $\mathbf{k}_\|=(k_m, k_y)$ and the slab thickness $L^\text{NCS}$. We assume open boundary conditions at both surfaces of the slab. This leads to
\begin{widetext}
\begin{equation}
H_\text{BCS} = \frac{1}{2} \sum_{\mathbf{k}_\|, l} \left[ \Psi_{(\mathbf{k}_\|,l)}^\dagger\,
    H_0^\text{NCS}(\mathbf{k}_\|)\, \Psi_{(\mathbf{k}_\|,l)}
  + \Psi_{(\mathbf{k}_\|,l-1)}^\dagger\, H_1^\text{NCS}(\mathbf{k}_\|)\, \Psi_{(\mathbf{k}_\|,l)}
  + \Psi_{(\mathbf{k}_\|,l+1)}^\dagger\, H_1^\text{NCS}(\mathbf{k}_\|)^\dagger\,
    \Psi_{(\mathbf{k}_\|,l)} \right] ,
\label{eq:Hamiltonian_fieldfree}
\end{equation}
with
\begin{align}
H_0^\text{NCS}(\mathbf{k}_\|) &=
\begin{pmatrix}
 -\mu -2 t \cos k_y  & \lambda  \sin k_y  & -\Delta^t \sin k_y  & \Delta^s\\
 \lambda  \sin k_y  & -\mu -2 t \cos k_y  & -\Delta^s & \Delta^t \sin k_y  \\
 -\Delta^t \sin k_y  & -\Delta^s & \mu +2 t \cos k_y  & \lambda  \sin k_y  \\
 \Delta^s & \Delta^t \sin  k_y  & \lambda  \sin k_y  & \mu + 2 t \cos k_y  \\
\end{pmatrix} , \\
H_1^\text{NCS}(\mathbf{k}_\|) &=
\begin{pmatrix}
-2 t \cos \frac{k_m}{\sqrt{2}} & \frac{\lambda}{2}\, e^{\frac{i k_m}{\sqrt{2}}} & -\frac{\Delta^t}{2}\, e^{\frac{i k_m}{\sqrt{2}}} & 0 \\
 -\frac{\lambda}{2}\, e^{\frac{i k_m}{\sqrt{2}}} & -2 t \cos \frac{k_m}{\sqrt{2}} & 0 & -\frac{\Delta^t}{2}\, e^{\frac{i k_m}{\sqrt{2}}} \\
 \frac{\Delta^t}{2}\, e^{\frac{i k_m}{\sqrt{2}}} & 0 & 2 t \cos \frac{k_m}{\sqrt{2}} & -\frac{\lambda}{2}\, e^{\frac{i k_m}{\sqrt{2}}} \\
 0 & \frac{\Delta^t}{2}\, e^{\frac{i k_m}{\sqrt{2}}} & \frac{\lambda}{2}\, e^{\frac{i k_m}{\sqrt{2}}} & 2 t \cos \frac{k_m}{\sqrt{2}}
\end{pmatrix} ,
\end{align}
and the spinors
\begin{align}
\Psi_{(\mathbf{k}_\|,l)} &= (c_{(\mathbf{k}_\|,l),\uparrow},c_{(\mathbf{k}_\|,l),\downarrow},
c_{(-\mathbf{k}_\|,l),\uparrow}^\dagger,c_{(-\mathbf{k}_\|,l),\downarrow}^\dagger)^T.
\end{align}
Next, we write the Hamiltonian in matrix form
\begin{equation}\label{eq:H_slab_NCS}
H_\text{BCS}=\frac{1}{2}\sum_{\mathbf{k}_\|} (\Psi^\text{NCS}_{\mathbf{k}_\|})^\dagger
\underbrace{
\begin{pNiceMatrix}
H_0^\text{NCS} (\mathbf{k}_\|)   & H_1^\text{NCS}(\mathbf{k}_\|)    & 0      & \Ldots & 0 \\
(H_1^\text{NCS})^\dagger(\mathbf{k}_\|)    & \Ddots & \Ddots & \Ddots & \Vdots\\
0      & \Ddots &        &        &  0\\
\Vdots & \Ddots &        &        & H_1^\text{NCS}(\mathbf{k}_\|)\\
0      & \Ldots & 0      & (H_1^\text{NCS})^\dagger(\mathbf{k}_\|)    & H_0(\mathbf{k}_\|)
\end{pNiceMatrix}}_{\mathcal{H}_\text{slab}^\text{NCS}(\mathbf{k}_\|)}
\Psi_{\mathbf{k}_\|}^\text{NCS} ,
\end{equation}
where the spinors now have $4L$ entries, $\Psi_{\mathbf{k}_\|}^\text{NCS}=(\Psi_{(\mathbf{k}_\|,l=0)}^T, \dots, \Psi_{(\mathbf{k}_\|,l=L^\text{NCS}-1)}^T)^T$, instead of four. This defines the BdG Hamiltonian $\mathcal{H}_\text{slab}^\text{NCS}(\mathbf{k}_\|)$ of a superconducting slab with (101) surfaces without a FI on top.

For the FI, we perform the same transformation on a BdG Hamiltonian with vanishing gap matrix and a normal-state Hamiltonian given in Eq.~\eqref{eq:normal_state_FI}. The FI slab is thus described by the Hamiltonian
\begin{equation}\label{eq:H_slab_FI}
H_\text{slab}^\text{FI}=\frac{1}{2}\sum_{\mathbf{k}_\|} (\Psi^\text{FI}_{\mathbf{k}_\|})^\dagger
\underbrace{
\begin{pNiceMatrix}
H_0^\text{FI} (\mathbf{k}_\|)   & H_1^\text{FI}(\mathbf{k}_\|)    & 0      & \Ldots & 0 \\
(H_1^\text{FI})^\dagger(\mathbf{k}_\|)    & \Ddots & \Ddots & \Ddots & \Vdots\\
0      & \Ddots &        &        &  0\\
\Vdots & \Ddots &        &        & H_1^\text{FI}(\mathbf{k}_\|)\\
0      & \Ldots & 0      & (H_1^\text{FI})^\dagger(\mathbf{k}_\|)    & H_0(\mathbf{k}_\|)
\end{pNiceMatrix}}_{\mathcal{H}_\text{slab}^\text{FI}(\mathbf{k}_\|)}
\Psi_{\mathbf{k}_\|}^\text{FI} ,
\end{equation}
where the spinor is $\Psi_{\mathbf{k}_\|}^\text{FI}=(\Psi_{(\mathbf{k}_\|,l=-L^\text{FI})}^T, \dots, \Psi_{(\mathbf{k}_\|,l=-1)}^T)^T$, with the thickness $L^\text{FI}$, and the blocks are given by
\begin{equation}
H_0^\text{FI} (\mathbf{k}_\|)=\begin{pmatrix}
 -\mu -2 t \cos k_y+V +h^z & h^x+i h^y  & 0  & 0\\
 h^x-i h^y  & -\mu -2 t \cos k_y-h^z+V  & 0 & 0  \\
0 & 0 & \mu +2 t \cos k_y - h^z -V & -h^x+i h^y  \\
0 & 0 & -h^x-i h^y  & \mu + 2 t \cos k_y + h^z-V  \\
\end{pmatrix}
\end{equation}
\end{widetext}
and
\begin{align}
H_1^\text{FI} (\mathbf{k}_\|)&= 2 t \cos\left(\frac{k_m}{\sqrt{2}}\right)\text{diag}\left(-1, -1,1 ,1 \right) .
\end{align}
We now combine Eqs.~\eqref{eq:H_slab_NCS} and~\eqref{eq:H_slab_FI}, leading to the BdG Hamiltonian of the NCS-FI heterostructure,
\begin{equation}\label{eq:H_slab_heterostructure}
\mathcal{H}_\text{slabs}(\mathbf{k}_\|)=\begin{pmatrix}
\mathcal{H}_\text{slab}^\text{FI}(\mathbf{k}_\|)& B(\mathbf{k}_\|)\\
B^\dagger(\mathbf{k}_\|)&\mathcal{H}_\text{slab}^\text{NCS}(\mathbf{k}_\|)
\end{pmatrix},
\end{equation}
where
\begin{equation}
B(\mathbf{k}_\|)=\begin{pNiceMatrix}
0 & 0 & \Ldots & 0      \\
\vdots & \Vdots & \Ddots& \Vdots\\
0 & 0 & \Ldots & 0\\
H_1^\text{FI}(\mathbf{k}_\|) &0 & \cdots &0
\end{pNiceMatrix},
\end{equation}
i.e., the off-diagonal blocks are empty except for blocks $H_1^\text{FI}(\mathbf{k}_\|)$ and $(H_1^\text{FI})^\dagger(\mathbf{k}_\|)$ in the top right and bottom left corner, respectively, and the spinor is given by $\Psi_{\mathbf{k}_\|}=(\Psi_{(\mathbf{k}_\|,l=-L^\text{FI})}^T, \dots, \Psi_{(\mathbf{k}_\|,l=L^\text{NCS}-1)}^T)^T$.
The Hamiltonian in Eq.~\eqref{eq:H_slab_heterostructure} is used to derive the transfer-matrix method in Sec.~\ref{sec:approximation}. Also, the results in Sec.~\ref{sec:approximation} are compared to the eigenstates of this Hamiltonian. Since we only need the surface states and the corresponding eigenvalues, we have used an implicitly restarted Lanczos method implemented in the SciPy function scipy.sparse.linalg.eigsh.

\section{Calculation of the wave function with the transfer-matrix method}
\label{sec:appendix_2}

The eigenvalues $\tau_j^{\text{NCS}}$ of the transfer matrix $T^\text{NCS}$ are determined by the eigenvalue equation
\begin{equation}\label{eq:eigvals_transfermatrix}
T^\text{NCS} t_j^\text{NCS} = \tau_j^{\text{NCS}}t_j^\text{NCS} ,
\end{equation}
which for $\tau_j^{\text{NCS}}\neq 0$ is equivalent to
\begin{align}
0&=\left[(\tau_j^{\text{NCS}})^{-1} (H_1^\text{NCS})^\dagger  + (H_0^\text{NCS}- E \boldone_4) \right.\notag \\
 &~~~~~\left. +\tau_j^{\text{NCS}} H_1^\text{NCS} \right]t^\text{NCS}_{j,u},
\end{align}
where we have defined $t_{j,u}^\text{NCS}$ to consist of the first four components of $t_j^\text{NCS}$, and the remaining four components follow as $t_{j,d}^\text{NCS}=\tau_j^{\text{NCS}} t_{j,u}^\text{NCS}$.

For numerical calculations, it is most efficient to solve the eigenvalue equation \eqref{eq:eigvals_transfermatrix} numerically. However, to find an analytic relationship between the eigenvalues $\tau_j^{\text{NCS}}$, we can use the fact that Eq.~\eqref{eq:eigvals_transfermatrix} is equivalent to finding the roots of an eighth-order polynomial
\begin{equation}\label{eq:tau_polynomial}
p(\tau)=\sum_{i=0}^{8} c_i \tau^i
\end{equation}
in $\tau_j^{\text{NCS}}$
with coefficients
\begin{widetext}
\begin{align}
c_8 = c_0^\ast &= \frac{1}{16}\, e^{2 \sqrt{2} i  k_m} \left[\left(\Delta^t\right)^2
  + \lambda ^2\right]^2+16 t^4 \cos ^4 \frac{k_m}{\sqrt{2}}
  - \frac{1}{2} \left(1+e^{i \sqrt{2} k_m}\right)^2 t^2 \left[\left(\Delta^t\right)^2-\lambda ^2
  \right], \\
c_7 = c_1^\ast &= 2 t \cos \frac{k_m}{\sqrt{2}}
   \left(8 t^2 \left( \cos \sqrt{2}k_m + 1 \right) (2 t \cos k_y+\mu )+e^{i \sqrt{2} k_m}
   \left\lbrace \left[\lambda ^2-\left(\Delta^t\right)^2\right] (2 t \cos k_y+\mu )-2 \Delta^s\Delta^t  \lambda \right\rbrace \right) , \\
c_6 = c_2^\ast &= 2 t^2 \left( \cos \sqrt{2} k_m + 1 \right) \left\lbrace 2 \left(\Delta^s\right)^2
   - 2 E_{\mathbf{k}_\|}^2+\left[\left(\Delta^t\right)^2-\lambda ^2\right] \left(3-2 \cos ^2k_y \right)
   + 6 (2 t \cos k_y+\mu )^2\right\rbrace \notag \\
   &\quad{} + \frac{1}{4} e^{i \sqrt{2} k_m} \left( \left[\left(\Delta^t\right)^2+\lambda ^2\right]
   \left\lbrace 2 E_{\mathbf{k}_\|}^2-\left[\left(\Delta^t\right)^2+\lambda
   ^2\right] \left(3-2 \cos^2 k_y\right)\right\rbrace\right. \notag \\
   &\qquad{} + \left. 2 \left[\left(\Delta^t\right)^2-\lambda ^2\right] \left[\left(\Delta^s\right)^2-(2 t \cos k_y+\mu )^2\right]-8 \Delta^s \Delta^t \lambda  (2 t \cos k_y+\mu )\right) \notag \\
   &\quad{} + 16 t^4 \left( \cos \sqrt{2} k_m + 1 \right)^2
   - \left(1+e^{i \sqrt{2} k_m}\right)^2 t^2
   \left[\left(\Delta^t\right)^2-\lambda ^2\right] , \\
c_5 = c_3^\ast &= 4 t \cos \frac{k_m}{\sqrt{2}} \bigg( (2 t \cos k_y+\mu ) \left\lbrace 2
   \left(\Delta^s\right)^2 - 2 E_{\mathbf{k}_\|}^2+\left[\left(\Delta^t\right)^2-\lambda ^2\right]
   \left(3-2 \cos^2 k_y\right) + 12 t^2\right\rbrace \notag \\
   &\qquad{} + \Delta^s \Delta^t \lambda  \left(6-4 \cos ^2 k_y\right)
   + 2 (2 t \cos k_y+\mu )^3 +12 t^2
   \cos \sqrt{2} k_m\, (2 t \cos k_y+\mu ) \notag \\
   &\qquad{} - e^{i \sqrt{2} k_m} \left\lbrace \Delta^s\Delta^t \lambda -\frac{1}{2}\left[\lambda ^2-\left(\Delta^t\right)^2\right] (2 t \cos k_y+\mu )\right\rbrace \bigg) , \\
c_4 &= \frac{1}{8} \left[\left(\Delta^t\right)^2+\lambda ^2\right]^2
  + \left\lbrace E_{\mathbf{k}_\|}^2 - \left(\Delta^s\right)^2-8 t^2
  \cos^2 \frac{k_m}{\sqrt{2}}
  - \left[\left(\Delta^t\right)^2+\lambda ^2\right] \left(\frac{3}{2}-\cos ^2 k_y\right)-\left( \mu
  + 2 t \cos k_y\right)^2\right\rbrace^2 \notag \\
&\quad{} - 4 \left(\frac{3}{2}-\cos^2k_y\right)
  \left\lbrace 8 \lambda ^2 t^2 \cos^2 \frac{k_m}{\sqrt{2}}
  + \left[\Delta^s\Delta^t -\lambda \left( \mu +2 t \cos k_y\right)\right]^2\right\rbrace  \notag \\
&\quad{} + 4 t^2 \cos^2\frac{k_m}{\sqrt{2}} \left\lbrace\left(\Delta^t\right)^2-\lambda^2
  + 2 \cos^2 \frac{k_m}{\sqrt{2}}
  \left[\lambda ^2-\left(\Delta^t\right)^2+4 t^2\right]+8 (\mu+2 t \cos k_y )^2\right\rbrace.
\end{align}
\end{widetext}
Due to the relation
\begin{equation}
c_i = c_{8-i}^\ast
\end{equation}
between the coefficients, if $\tau_j^{\text{NCS}}$ is an eigenvalue of the transfer matrix, then we find another eigenvalue, which we will call $\tau_{j+4}^{\text{NCS}}$ as
\begin{equation}
\tau_{j+4}^{\text{NCS}}=\frac{1}{(\tau_j^{\text{NCS}})^\ast}.
\end{equation}

Using the eigenvalues $\tau_j^\text{NCS}$ of the transfer matrix, we now define the quantity $k_l^{(j)} =-\sqrt{2}i\ln \tau_j^\text{NCS}$. With this definition, we can rewrite Eq.~\eqref{eq:eigvals_transfermatrix} as
\begin{equation}\label{eq:transfermatrix_hamiltonian}
[\mathcal{H}(\mathbf{k}_\|,k_l^{(j)})-E_{\mathbf{k}_\|}]\,t_{j,u}^\text{NCS} = 0.
\end{equation}
Finding the eigenvalue of the transfer matrix is therefore equivalent to finding the complex solution $k_l^{(j)}$ of
\begin{equation}
0 = \det[\mathcal{H}(\mathbf{k}_\|,k_l^{(j)})-E_{\mathbf{k}_\|}],
\end{equation}
which is equivalent to solving the two equations
\begin{equation}\label{eq:kl_equation}
E_{\mathbf{k}_\|}^2-(\xi^\pm_{\mathbf{k}_\|,k_l})^2-(\Delta^\pm_{\mathbf{k}_\|,k_l})^2 = 0.
\end{equation}

As there is a one-to-one correspondence between the solutions $k_l^{(j)}$ of this equation and the roots $\tau_j^\text{NCS}$ of Eq.~\eqref{eq:tau_polynomial}, we find eight complex solutions, which occur in pairs with $\text{Re}(k_l^{(j+4)}) = \text{Re}(k_l^{(j)})$ and $\text{Im}(k_l^{(j+4)}) = -\text{Im}(k_l^{(j)})$.
 Using Eq.~\eqref{eq:wavefunction_transfermatrix}, we see that the surface state is a linear combination of the eigenvectors of the transfer matrix with prefactors that either decay (for $\tau_j^\text{NCS}<1$) or increase (for  $\tau_j^\text{NCS}>1$) exponentially in $l$. Since the wave function $\psi$ has to be bounded, the coefficients $\alpha_j^{\text{NCS}}$ which correspond to terms with exponential increase, have to vanish. This is the cases for eigenvalues $\tau_j^\text{NCS}>1$, or equivalently values of $k_l^{(j)}$ with negative real part. If any of the solutions $k_l^{(j)}$ were real, the surface state energy $E_{\mathbf{k}_\|}$ would be an eigenvalue of the BdG Hamiltonian in Eq.~\eqref{eq:bdg_hamiltonian} for $\mathbf{k}=(\mathbf{k}_\|,k_l^{(j)})$. However, since we want to find a surface state in the bulk gap we consider only energies $E_{\mathbf{k}_\|}<\Delta_{\mathbf{k}}^\pm$ such that none of the $k_l^{(j)}$ are real. We can therefore ignore the solutions with negative real part, which we choose to be $k_l^{(5)}$, \dots, $k_l^{(8)}$, and only proceed with the solutions $k_l^{(1)}$, \dots, $k_l^{(4)}$ with positive real part. The corresponding eigenvectors $t_{j,u}^{\text{NCS}}$ follow from Eq.\eqref{eq:transfermatrix_hamiltonian} as
\begin{align}\label{eq:eigenvector_TNCS}
t_{j,u}^{\text{NCS}} = \begin{pmatrix}
u^\pm_{\mathbf{k}^{(j)}}\\
\gamma^\pm_{\mathbf{k}^{(j)}}\, i \sigma^y\, u^\pm_{\mathbf{k}^{(j)}}
\end{pmatrix},
\end{align}
where we introduced the notation $\mathbf{k}^{(j)}=(\mathbf{k}_\|, k_l^{(j)})$ and $u^\pm_{\mathbf{k}^{(j)}}$ are the eigenvectors of the normal-state Hamiltonian given in Eq.~\eqref{eq:eigvec_normalstatehamiltonian}. The prefactor in the lower components in Eq.\ \eqref{eq:eigenvector_TNCS} is
\begin{equation}\label{eq:gamma}
\gamma^\pm_{\mathbf{k}^{(j)}}=\frac{\xi^\pm_{\mathbf{k}^{(j)}}-E_{\mathbf{k}_\|}}{\Delta^\pm_{\mathbf{k}^{(j)}}}.
\end{equation}
The sign $\pm$ has to be chosen according to the sign for which Eq.~\eqref{eq:kl_equation} is satisfied.

In analogy to the NCS case, the eigenvalues $\tau_j^\text{FI}$ for the FI can be used to define $\kappa_\perp^{(j)}=\sqrt{2} \ln \tau_j^\text{FI}$. We then find that $\kappa_\perp^{(j)}$ are the solutions of $\det(\mathcal{H}^\text{FI}(\mathbf{k}_\|, k_l=i \kappa_\perp)-E_{\mathbf{k}_\|})=0$ with
\begin{equation}
\mathcal{H}^\text{FI}(\mathbf{k}_\|, k_l)= \begin{pmatrix}
h_\text{FI}(\mathbf{k}_\|, k_l) & 0 \\
0 & -[h_\text{FI}(-\mathbf{k}_\|, -k_l)]^T
\end{pmatrix},
\end{equation}
which are real. In order to ensure that $\psi(\mathbf{k}_\|)$ is bounded, we again have to set the prefactors $\alpha_j^\text{FI}$ in Eq.~\eqref{eq:wavefunction_transfermatrix} to zero for terms which increase exponentially for $l\rightarrow -\infty$.  We can therefore restrict ourselves to positive solutions $\kappa_\perp^{(j)}$. We find four positive real solutions
\begin{equation}\label{eq:kappa}
\kappa_\perp^{(j)} \equiv \kappa_\perp^{\zeta, \sigma}
= \sqrt{2}~\text{arcosh} \frac{V-\mu-2 t \cos k_y -\zeta E_{\mathbf{k}_\|}\! + \sigma |\mathbf{h}|}{4 t \cos \frac{k_m}{\sqrt{2}}} .
\end{equation}
The first four components of the corresponding eigenvectors of the transfer matrix span the nullspace of $\mathcal{H}^\text{FI}(\mathbf{k}_\|, k_l=i \kappa_\perp)-E_{\mathbf{k}_\|}$. If we rewrite the magnetization $\mathbf{h}$ in spherical coordinates,
\begin{equation}
\mathbf{h}=\begin{pmatrix}
h^x\\ h^y\\h^z
\end{pmatrix}= h \begin{pmatrix}
\cos \eta \sin \xi \\ \sin\eta \sin \xi \\ \cos \xi
\end{pmatrix},
\end{equation}
they take the simple form
\begin{align}
t_{\zeta=1,\sigma=1, u}^\text{FI}&=\left(e^{- i \eta} \cos(\xi/2), \sin(\xi/2),0,0\right)^T,\\
t_{\zeta=1,\sigma=-1,u}^\text{FI}&=\left(-e^{- i \eta} \sin(\xi/2), \cos(\xi/2),0,0\right)^T,\\
t_{\zeta=-1,\sigma=1,u}^\text{FI}&=\left(0,0,e^{i \eta} \cos(\xi/2), \sin(\xi/2)\right)^T,\\
t_{\zeta=-1,\sigma=-1,u}^\text{FI}&=\left(0,0,-e^{i \eta} \sin(\xi/2), \cos(\xi/2)\right)^T.
\end{align}
Next, we examine the boundary conditions in Eqs.~\eqref{eq:boundary_0} and~\eqref{eq:boundary_1}, which are two vector equations of length four, i.e., eight scalar equations. These equations are linear in the coefficients $\alpha_1^{\text{NCS}}$, \dots, $\alpha_4^{\text{NCS}}$, and $\alpha_{\zeta,\sigma}^{\text{FI}}$ and can thus be written as
\begin{equation} \label{eq:nullspace}
0=M \left(\alpha_1^{\text{NCS}}, \dots, \alpha_4^{\text{NCS}},\alpha_{1,1}^{\text{FI}},\alpha_{1,-1}^{\text{FI}},\alpha_{-1,1}^{\text{FI}},\alpha_{-1,-1}^{\text{FI}}\right)^T ,
\end{equation}
with an $8\times 8$ matrix $M$ which is given by
\begin{align}\label{eq:M}
M=\begin{pmatrix}
M_1&M_2\\
M_3&M_4
\end{pmatrix}
\end{align}
with the blocks
\begin{align}
M_1&=\begin{pmatrix}
 1 & 1 & 1 & 1 \\
 q_1 & q_2 & q_3 & q_4 \\
 \gamma_1 q_1 & \gamma_2 q_2 & \gamma_3 q_3 & \gamma_4 q_4 \\
 -\gamma_1 & -\gamma_2 & -\gamma_3 & -\gamma_4 \\
\end{pmatrix} , \\
M_2 &= \begin{pmatrix}
 -e^{-i \eta } \cos \frac{\xi }{2} & e^{-i \eta } \sin \frac{\xi }{2} & 0 & 0 \\
 -\sin \frac{\xi }{2} & -\cos \frac{\xi }{2} & 0 & 0 \\
 0 & 0 & -e^{i \eta } \cos \frac{\xi }{2} & e^{i \eta } \sin \frac{\xi }{2} \\
 0 & 0 & -\sin \frac{\xi }{2} & -\cos \frac{\xi }{2} \\
\end{pmatrix} , \\
M_3 &=
\begin{pmatrix}
  m_{1,1} & m_{1,2} & m_{1,3} & m_{1,4} \\
  m_{2,1} & m_{2,2} & m_{2,3} & m_{2,4} \\
  m_{3,1} & m_{3,2} & m_{3,3} & m_{3,4} \\
  m_{4,1} & m_{4,2} & m_{4,3} & m_{4,4}
\end{pmatrix} ,
\end{align}
and
\begin{widetext}
\begin{align}
M_4&=2t \cos \left(\frac{k_m}{\sqrt{2}}\right)\begin{pmatrix}
e^{-\frac{\kappa_\perp^{1,1}}{\sqrt{2}}-i \eta } \cos \frac{\xi }{2} &
-e^{-\frac{\kappa_\perp^{1,-1}}{\sqrt{2}}-i \eta } \sin \frac{\xi }{2} & 0 & 0 \\
 e^{-\frac{ \kappa_\perp^{1,1}}{\sqrt{2}}} \sin \frac{\xi }{2} & e^{-\frac{\kappa_\perp^{1,-1}}{\sqrt{2}}} \cos \frac{\xi }{2} & 0 & 0 \\
 0 & 0 & -e^{-\frac{ \kappa_\perp^{-1,1}}{\sqrt{2}}+i \eta } \cos \frac{\xi }{2} & e^{-\frac{\kappa_\perp^{-1,-1}}{\sqrt{2}}+i \eta } \sin \frac{\xi }{2} \\
 0 & 0 & -e^{-\frac{\kappa_\perp^{-1,1}}{\sqrt{2}}} \sin \frac{\xi }{2} & -e^{-\frac{ \kappa_\perp^{-1,-1}}{\sqrt{2}}} \cos \frac{\xi }{2} \\
\end{pmatrix},
\end{align}
\end{widetext}
where \begin{align}
q_j &= \pm \frac{l^x_{\mathbf{k}_\|, k_l^{(j)}}+i l^y_{\mathbf{k}_\|, k_l^{(j)}}}{\sqrt{\left(l^x_{\mathbf{k}_\|, k_l^{(j)}}\right)^2+\left(l^y_{\mathbf{k}_\|, k_l^{(j)}}\right)^2}},\\
m_{1,j} &= \frac{1}{2} e^{-i\frac{k_l^{(j)}}{\sqrt{2}}} \left[-4 t \cos \frac{k_m}{\sqrt{2}}
  +e^{-\frac{i k_m}{\sqrt{2}}} q_j (\gamma_j \Delta^t-\lambda )\right],\\
m_{2,j} &= -\frac{1}{2}e^{-i\frac{ k_l^{(j)}}{\sqrt{2}}} \left[4 q_j t \cos \frac{k_m}{\sqrt{2}}
  +e^{-\frac{i k_m}{\sqrt{2}}} (\gamma_j \Delta^t-\lambda )\right] ,\\
m_{3,j}&=\frac{1}{2}e^{-i\frac{ k_l^{(j)}}{\sqrt{2}}}  \left[4 \gamma_j q_j t \cos \frac{k_m}{\sqrt{2}}
  -e^{-\frac{i k_m}{\sqrt{2}}} (\gamma_j \lambda +\Delta^t)\right], \\
m_{4,j}&=-\frac{1}{2} e^{-i\frac{k_l^{(j)}}{\sqrt{2}}} \notag \\
&\quad{} \times\left[4 \gamma_j t \cos \frac{k_m}{\sqrt{2}}
  +e^{-\frac{i k_m}{\sqrt{2}}} q_j(\gamma_j \lambda +\Delta^t)\right].
\end{align}
Nontrivial solutions require $\det M=0$. For $\mathbf{h}=0$, this is achieved for $E_{\mathbf{k}_\|}=0$, as expected. The derivation is relegated to Appendix~\ref{sec:appendix_3}.
For $\mathbf{h}\neq 0$, the time-reversal-symmetry breaking term introduces a dispersion to the previously flat band and Eq.~\eqref{eq:nullspace} has to be solved numerically.

Finally, the surface state can be obtained as
\begin{equation}
\label{eq:psi_transfer}
\psi_l({\mathbf{k}_\|}) = \begin{cases}
\displaystyle\frac{1}{n_{\mathbf{k}_\|}}\exp\left(i \phi_{\mathbf{k}_\|}\right) \sum_{j=1}^{4} \alpha_j^\text{NCS}({\mathbf{k}_\|}) \\
\displaystyle\;{} \times\exp\left({i k_l^{(j)}({\mathbf{k}_\|}) \frac{l}{\sqrt{2}}}\right) t_{j,u}^\text{NCS}({\mathbf{k}_\|}) \quad\text{for }l \geq 0,\\
\displaystyle \frac{1}{n_{\mathbf{k}_\|}}\exp\left(i \phi_{\mathbf{k}_\|}\right)
\sum_{\zeta,\sigma \in \lbrace+1,-1\rbrace } \alpha_{\zeta,\sigma}^\text{FI}({\mathbf{k}_\|}) \\
\displaystyle\;{} \times \exp\left({\kappa_\perp^{\zeta,\sigma}({\mathbf{k}_\|}) \frac{l}{\sqrt{2}}}\right) t_{\zeta,\sigma,u}^\text{FI}({\mathbf{k}_\|}) \quad\text{for }l<0,
\end{cases}
\end{equation}
with the normalization constant
\begin{align}
n_{\mathbf{k}_\|} &= \left[\sum_{\zeta,\sigma}
\frac{|\alpha_{\zeta,\sigma}^\text{FI}|^2}{\exp\left(\sqrt{2} \kappa_\perp^{\zeta,\sigma}\right)-1}  \right. \notag\\
&\quad{} + \left.
\sum_{m,n=1}^4\frac{\left(\alpha_{m}^\text{NCS}\right)^\ast \alpha_{n}^\text{NCS} (t_{m,u}^\text{NCS})^\dagger t_{n,u}^\text{NCS} }{1-\exp\left\lbrace\sqrt{2}i\left[k_l^{(n)}-\left(k_l^{(m)}\right)^\ast\right]\right\rbrace}
\right]^{1/2}
\end{align}
and a phase
\begin{equation}
\phi_{\mathbf{k}_\|} = \arg\left( \sum_{j=1}^{4} \alpha_j^\text{NCS}({\mathbf{k}_\|})^\ast  ~ [t_{j,u}^\text{NCS}({\mathbf{k}_\|})]^\dagger (1,i,0,0)^T
\right) ,
\end{equation}
which ensures the mirror symmetry in the $y$ direction.

\section{Surface states for \texorpdfstring{$\mathbf{h}=0$}{h=0}}
\label{sec:appendix_3}

For our model system with $C_4$ point group at $\mathbf{h}=0$, the surface state obtained from the transfer-matrix method can be calculated analytically. Finding the transfer matrix requires us to determine the energy $E_{\mathbf{k}_\|}$, for which Eqs.~\eqref{eq:boundary_0} and \eqref{eq:boundary_1} allow nontrivial solutions for the coefficients $\alpha^\text{NCS}_j$, $\alpha^\text{FI}_j$. These equations mean that the coefficients $\alpha^\text{NCS}_j$, $\alpha^\text{FI}_j$ are given by the nullspace of the matrix $M$ given in Eq.~\eqref{eq:M}, where the values for $ \kappa_\perp^{\zeta,\sigma}$ and $k_l^{(j)}$ have to be determined as described in Appendix~\ref{sec:appendix_2} to be consistent with the value of $E_{\mathbf{k}_\|}$. If $\mathbf{h}=0$, then for $E_{\mathbf{k}_\|}=0$, Eq.~\eqref{eq:kappa} implies $\kappa \equiv \kappa_\perp^{1,1}=\kappa_\perp^{1,-1}= \kappa_\perp^{-1,1}=\kappa_\perp^{-1,-1}$.

Using Eqs.~\eqref{eq:kl_equation} and \eqref{eq:gamma}, we also find that $\gamma_{\mathbf{k}_\|,k_l^{(j)}}^\pm \in\lbrace i,-i\rbrace$. In particular, labeling the positive-helicity solutions, which satisfy Eq.\ \eqref{eq:kl_equation} with a positive sign, as $k_l^{(1)}$ and $k_l^{(2)}$ with $\text{Re}(k_l^{(1)})>\text{Re}(k_l^{(2)})$ and the negative-helicity solutions which satisfy Eq.\ \eqref{eq:kl_equation} with a negative sign as $k_l^{(3)}$ and $k_l^{(4)}$, we find $\gamma_1=-\gamma_2=-\gamma_3=-\gamma_4 = i$ within the region $\mathcal{F}_r$ \cite{SBT12}. A relabeling would change which of the $\gamma_j$ has a different sign than the others and for a surface momentum in the region $\mathcal{F}_l$, all $\gamma_j$ switch sign.
With these simplifications for $ \kappa_\perp^{\zeta,\sigma}$ and $\gamma_j$, one can easily check that $\det M=0$ so that the nullspace is nontrivial, which proves that there is a surface state at ${E_{\mathbf{k}_\|}=0}$. We can also calculate the nullspace, which leads to
\begin{widetext}\begin{align}
\alpha^\text{NCS}_1 &= 0 , \\
\alpha^\text{NCS}_2 &= \left[16 t^2 \cos ^2\left(\frac{k_m}{\sqrt{2}}\right) (q_3-q_4)
  \left(e^{\kappa/\sqrt{2}} + e^{\frac{-\kappa + i  \left(k_l^{(3)}+k_l^{(4)}\right)}{\sqrt{2}}}-e^{i k_l^{(3)}/\sqrt{2}} - e^{i k_l^{(4)}/\sqrt{2}}\right) \right. \notag \\
&{} - \left. 2 i t \left(1+e^{-i \sqrt{2} k_m}\right) \left(\Delta^t-i \lambda \right) \left(e^{i k_l^{(3)}/\sqrt{2}} - e^{i k_l^{(4)}/\sqrt{2}}\right) (q_3 q_4+1)
   + \left(\Delta^t-i \lambda \right)^2 \left(-e^{\frac{\kappa -2 i
   k_m}{\sqrt{2}}}\right) (q_3-q_4)\right] e^{i\frac{ k_l^{(2)}-k_l^{(4)}}{\sqrt{2}}} , \\
\alpha^\text{NCS}_3 &= \left[16 t^2 \cos ^2\left(\frac{k_m}{\sqrt{2}}\right) (q_4-q_2)
   \left(e^{\kappa/\sqrt{2}} + e^{\frac{-\kappa +i \left(k_l^{(2)}+k_l^{(4)}\right)}{\sqrt{2}}}
   - e^{i k_l^{(2)}/\sqrt{2}} - e^{i k_l^{(4)}/\sqrt{2}}\right) \right.\notag \\
&{} - \left. 2 i \left(1+e^{-i \sqrt{2} k_m}\right) t \left(\Delta^t-i \lambda \right)
   \left(e^{i k_l^{(4)}/\sqrt{2}} - e^{i k_l^{(2)}/\sqrt{2}}\right) (q_2 q_4+1)
   +\left(\Delta^t-i \lambda \right)^2 \left(-e^{\frac{\kappa -2 i
   k_m}{\sqrt{2}}}\right) (q_4-q_2)\right] e^{i\frac{k_l^{(3)}-k_l^{(4)}}{\sqrt{2}}} , \\
\alpha^\text{NCS}_4 &= 16 t^2 \cos^2\left(\frac{k_m}{\sqrt{2}}\right) (q_2-q_3)
   \left(e^{\kappa/\sqrt{2}}
   + e^{\frac{-\kappa +i \left(k_l^{(2)}+k_l^{(3)}\right)}{\sqrt{2}}}
   - e^{i k_l^{(2)}/\sqrt{2}} - e^{i k_l^{(2)}/\sqrt{2}}\right) \notag \\
&{} -2 i t \left(1+e^{-i \sqrt{2} k_m}\right)
    \left(\Delta^t-i \lambda \right) \left(e^{i k_l^{(2)}/\sqrt{2}}
    - e^{i k_l^{(3)}/\sqrt{2}}\right) (q_2 q_3+1)
    +(\Delta^t-i \lambda )^2 \left(-e^{\frac{\kappa -2 i k_m}{\sqrt{2}}}\right) (q_2-q_3)
\end{align}\end{widetext}
for the NCS components and
\begin{align}
&\left(\alpha^\text{FI}_{1,1},\alpha^\text{FI}_{1,-1},\alpha^\text{FI}_{-1,1},\alpha^\text{FI}_{-1,-1}\right)^T\notag \\
&=-M_2^{-1} M_1 \left(\alpha^\text{NCS}_1,\alpha^\text{NCS}_2,\alpha^\text{NCS}_3,\alpha^\text{NCS}_4\right)^T
\end{align}
for the FI components in the region $\mathcal{F}_r$. The nullspace in the region $\mathcal{F}_l$ can be computed analogously. This means that it is possible to calculate the zero-energy surface states of the system at $\mathbf{h}=0$ analytically if the values for $k_l^{(j)}$ which satisfy Eq.~\eqref{eq:kl_equation} are given.
This equation is equivalent to the eighth order polynomial in $\tau_j^\text{NCS}=\exp(i k_l^{(j)}/\sqrt{2})$ given by Eq.~\eqref{eq:tau_polynomial} and can typically not be solved analytically.

Note that one could also find a surface state for a heterostructure of finite thickness using the same transfer-matrix method, taking the boundary conditions \eqref{eq:top_surface} and \eqref{eq:bottom_surface} into account. In that case the negative solutions $\kappa_\perp^{(j)}$ and solutions $k_l^{(j)}$ with negative imaginary part are not neglected. Including Eqs.\ \eqref{eq:top_surface} and \eqref{eq:bottom_surface}, we then have a system of 16 homogeneous linear equations for the $\alpha^\text{NCS}_j$ and $\alpha^\text{FI}_j$. The solution would then be found by identifying the values of $E$ for which the the determinant of the coefficient matrix is zero and the resulting eigenvalue and eigenstate would be identical to the ones found by diagonalizing the slab Hamiltonian. In practice, this procedure is likely slower than using some numerical method to directly find the lowest-energy eigenvalue of the slab Hamiltonian, except for very large slab thicknesses, because it includes numerical root finding both for the determinant and to calculate the $k_l^{(j)}$. It does, however, have the advantage that it uses significantly less storage for the eigenvectors, which can be fully calculated with just the eight values of $\alpha^\text{NCS}_j$, $\alpha^\text{FI}_j$, $k_l^{(j)}$, and $\kappa_\perp^{\zeta,\sigma}$, while the full eigenvector contains $4(L^\text{FI}+L^\text{NCS})$ components.

\section{Numerical comparison of the transfer-matrix surface states with results from exact diagonalization}
\label{sec:appendix_4}

\begin{figure}[tbp]
\centering
\includegraphics[width=0.48\textwidth]{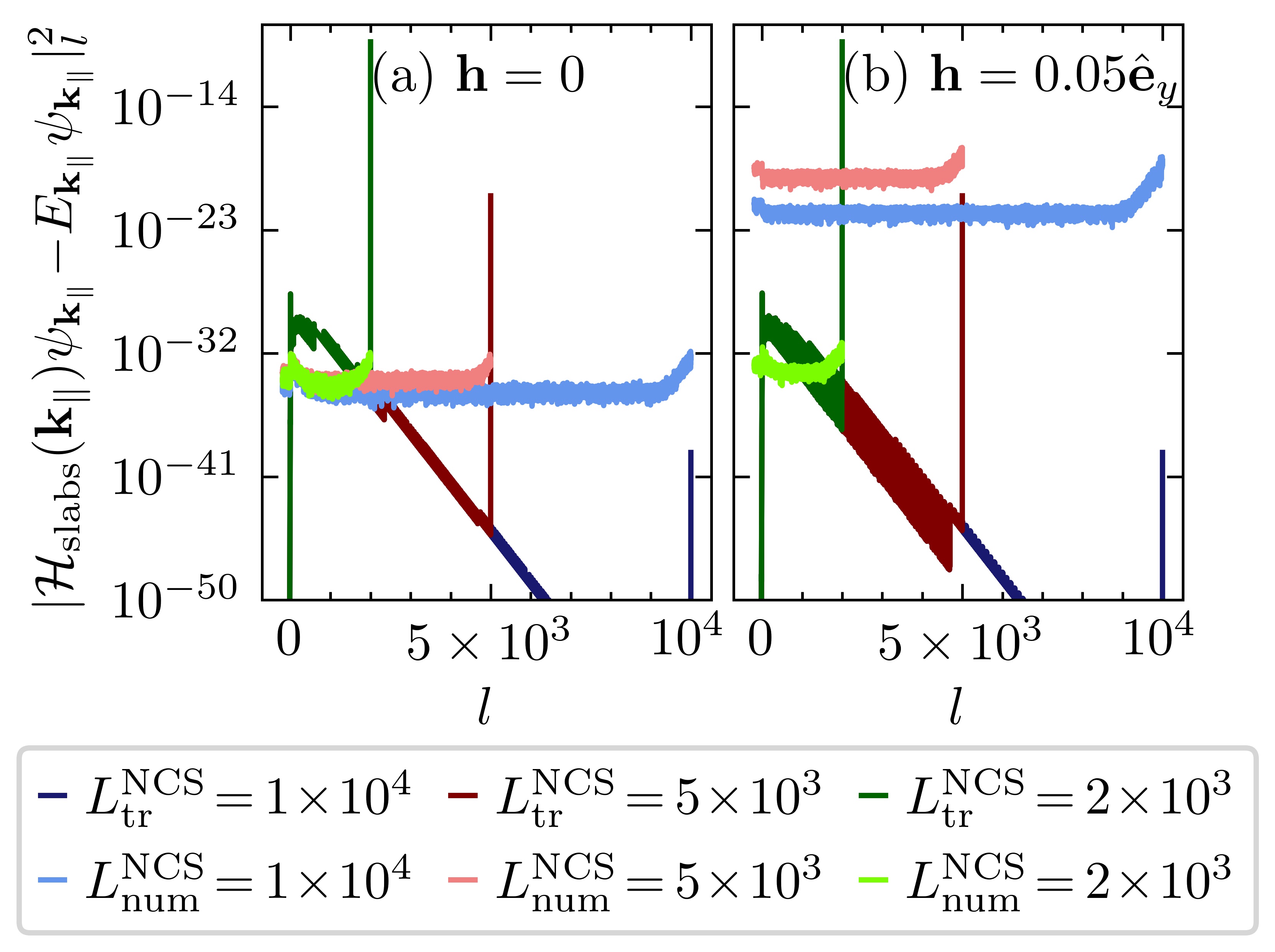}
\caption{ Comparison of the zero-energy surface state for $\mathbf{k}_\|$ inside the projections of the nodal lines of the infinite NCS-FI heterostructure to the state obtained from numerical diagonalization of the Hamiltonian of the finite slab. (a) Squared modulus $|(\mathcal{H}_\text{slabs}^{L^\text{NCS}}(\mathbf{k}_\|) - E_{\mathbf{k}_\|})\psi(\mathbf{k}_\|)|^2$ as a function of $l$ for the case of $\mathbf{h}=0$. (b) The same for $\mathbf{h}=0.05~\hat{\mathbf{e}}_y$. Here, results from the transfer-matrix method, cut off at different thicknesses $L^\text{NCS}$, are shown in darker colors, whereas results from exact diagonalization for the finite slab are shown in lighter colors.}
\label{fig:transfermatrix_eigenstates}
\end{figure}

Real systems are of finite thickness. In the following, we present numerical evidence that these results are a good approximation for the eigenstates of a sufficiently thick slab. Note that to this purpose, it is not helpful to numerically calculate the surface states of a finite slab and compare it---or its modulus---to the surface states obtained by the transfer-matrix method. Due to the hybridization of the surface states at the $l=0$ surface and at the $l=L^\text{NCS}$ surface, these results may not be close to each other even if the state obtained with the transfer-matrix method is an eigenstate of the Hamiltonian in the limit of infinitely thick slabs. Therefore, we instead show the squared modulus of the vector $(\mathcal{H}_\text{slabs}^{L^\text{NCS}}(\mathbf{k}_\|)-E_{\mathbf{k}_\|})\psi(\mathbf{k}_\|)$ in Fig.~\ref{fig:transfermatrix_eigenstates}. For an exact eigenstate of the Hamiltonian $\mathcal{H}_\text{slabs}^{L^\text{NCS}}(\mathbf{k}_\|)$, this quantity should vanish for all $l$.
For the numerical calculation, we use the parameter values $k_m=1$, $k_y=0.05$, $t=1$, $\mu=-4$, $\lambda=0.05$, $V=3.5$, $\Delta^s=0.04$, and $\Delta^t=0.05$. We choose $L^\text{FI}=200$ as the thickness of the ferromagnetic layer.

For comparison, we plot the results for eigenstates determined by exact diagonalization of the slab Hamiltonian for $L^\text{NCS}=2000$ (light green), $L^\text{NCS}=5000$ (light red), and $L^\text{NCS}=10000$ (light blue). Figure \ref{fig:transfermatrix_eigenstates}(a) shows the time-reversal-symmetric case $\mathbf{h}=0$, while Fig.\ \ref{fig:transfermatrix_eigenstates}(b) shows the results for a nonzero magnetization $\mathbf{h}=0.05\;\hat{\mathbf{e}}_y$ of the FI. We see that in all cases, the results are almost constant at about $10^{-20}$ to $10^{-35}$, which is expected due to the calculation at finite machine precision.

In darker colors, we show the squared modulus $|(\mathcal{H}_\text{slabs}^{L^\text{NCS}}(\mathbf{k}_\|) - E_{\mathbf{k}_\|})\psi(\mathbf{k}_\|)|^2$ for states $\psi$ obtained by the transfer-matrix method and cut off at the corresponding widths $L^\text{NCS}=2000, 5000, 10000$. See Eq.\ \eqref{eq:psi_transfer} in Appendix \ref{sec:appendix_2}.
 These plots show that for large values of $L^\text{NCS}$, the transfer-matrix method is a good approximation of the surface state, since the value of $|(\mathcal{H}_\text{slabs}^{L^\text{NCS}}(\mathbf{k}_\|)-E_{\mathbf{k}_\|})\psi(\mathbf{k}_\|)|^2$ is similar to or even lower than the result obtained by numerical diagonalization. For small slab thickness, e.g., for $L^\text{NCS}=2000$, the approximation fails at $l=L^\text{NCS}-1$, as evidenced by the dark green line that rises to a value of approximately $10^{-10}$, which is much higher than the value of approximately $10^{-32}$ obtained by numerical diagonalization. This is expected since the transfer matrix method neglects the boundary conditions at $l=L^\text{NCS}-1$ and $l=-L^\text{FI}$.

It can also be seen that due to the explicit representation of an exponential decay in Eq.~\eqref{eq:psi_transfer}, the transfer matrix is much better suited to calculate the surface state for large values of $l$, whereas exact diagonalization results are more limited by machine precision. Note, however, that none of the exceedingly small numbers in the plot are supposed to be measurable quantities. On the contrary, the only relevant feature for our purposes is that the results stay below machine precision, such that they are a good estimate for the surface state.

\section{Adiabaticity}
\label{sec:appendix_5}

Adiabaticity can easily be defined in the case that the region $\mathcal{F}$ does not include the entirety of $\mathcal{F}_l \cup \mathcal{F}_r$ so that the infimum of the energy of the first excited quasiparticle state is nonzero,
\begin{align}
\Delta &= \inf \left( \left\lbrace \Delta^\nu_{\mathbf{k}=(\mathbf{k}_\|, k_\perp)} \Big|\,
  \nu = \pm, \mathbf{k}_\|\in \mathcal{F}, k_\perp\in[-\pi,\pi) \right\rbrace \right)
  \nonumber \\
&> 0 .
\end{align}
In this case, the dynamics is certainly adiabatic if the ramp time $T_i$ is large compared to the timescale $\hbar/\Delta$ given by the inverse of this excitation energy, where we set $\hbar =1$.

However, if $\mathcal{F}$ gets arbitrarily close to the projections of the nodal lines the timescale $\hbar/\Delta_\mathrm{min}$ diverges and a different condition is required. We note that the excited quasiparticle states the dispersion of which approaches zero energy at the nodes are bulk states. In this case, we compare the density of states of the surface states and the local density of states of bulk states at the surface to assess the adiabaticity. In the field-free system, the density of surface states is a Dirac delta peak at $E=0$. Integrating over this peak in an interval $E\in [-\hbar/T_i,\hbar/T_i]$ yields a constant. The local density of bulk states close to $E=0$ is linear, i.e., $D(E)\propto |E|$. The local density of bulk states at the surface is also independent of the slab thickness if the thickness is sufficiently large. We can therefore choose a timescale $T_i$ that is large enough such that the (constant) number of surface states in the energy interval $[-\hbar/T_i,\hbar/T_i]$ is large compared to the number of bulk states, which is proportional to $\hbar/T_i$. Under this condition, we expect that the mixing of surface states with bulk states can be neglected so that it is still reasonable to assume that the system stays in the instantaneous eigenstates, i.e., the surface states.

\section{Boundaries of the region with zero-energy surface states}
\label{sec:appendix_6}

In this appendix, we will derive an expression for the boundaries of the region $\mathcal{F}_r$ that hosts the zero-energy surface states in the part of the sBZ with $k_m>0$, depicted in Fig.~\ref{fig:sketch_blobs}. The region $\mathcal{F}_l$ can be determined by reflecting $\mathcal{F}_r$ at the $k_y$-axis.
The boundaries of $\mathcal{F}_r$ are determined by projecting the bulk nodal lines onto the sBZ. Therefore, it can be found by eliminating the perpendicular momentum component $k_l$ from the equations
\begin{align}
0&=\epsilon_{\mathbf{k}_\|, k_l}-\lambda |\mathbf{l}_{\mathbf{k}_\|, k_l}| , \label{eq:fermi_surf}\\
0&=\Delta^s-\Delta^t |\mathbf{l}_{\mathbf{k}_\|, k_l}|. \label{eq:gap_zero}
\end{align}
Eq.~\eqref{eq:fermi_surf} defines the negative-helicity Fermi surface and Eq.~\eqref{eq:gap_zero} are the roots of the negative-helicity gap function. The intersections of the two surfaces described by these equations are the nodal lines of the gap.
Solving Eq.~\eqref{eq:gap_zero} yields
\begin{equation}
k_l=-k_m\pm\sqrt{2} \arcsin\left[\sqrt{\left(\frac{\Delta^s}{\Delta^t}\right)^2-\sin^2k_y} \right] .
\end{equation}
Substitution into Eq.~\eqref{eq:fermi_surf} gives the implicit equation
\begin{align}\label{eq:boundary_implicit}
0&=\frac{\Delta^s}{\Delta^t}\lambda+\mu+2 t \cos k_y+4 t \cos\left(\frac{k_m}{\sqrt{2}}\right)\notag\\
&\quad{}\times\left\lbrace \cos\frac{k_m}{\sqrt{2}}\: \sqrt{1-\left[\left(\frac{\Delta^s}{\Delta^t}\right)^2-\sin^2 k_y\right]} \right.\notag \\
&\qquad{} + \left. \sin\frac{k_m}{\sqrt{2}}\:
  \sqrt{\left(\frac{\Delta^s}{\Delta^t}\right)^2-\sin^2 k_y}\right\rbrace
\end{align}
for the boundary.
We see that this equation can only be satisfied if
\begin{equation}
\left(\frac{\Delta^s}{\Delta^t}\right)^{\!2} - \sin^2 k_y\in [0,1]
\end{equation}
because otherwise one of the square roots would be imaginary such that the entire term in Eq.~\eqref{eq:boundary_implicit} could not be zero.
This leads to the minimum and maximum values of $k_y$ as
\begin{equation}
k_y^\text{max}=-k_y^\text{min}=\arcsin\left|\frac{\Delta^s}{\Delta^t}\right|.
\end{equation}
Solving Eq.~\eqref{eq:boundary_implicit} for $k_m$ gives the solutions
\begin{widetext}
\begin{align}
k_m^\nu(k_y) &= \sqrt{2}\: \arccos \left\lbrace\frac{1}{2} \left[\left(\frac{\Delta^s}{\Delta^t}\right)^{\!2}
  -\sin^2 k_y\right]-\sqrt{1-\left[\left(\frac{\Delta^s}{\Delta^t}\right)^{\!2}
  -\sin^2 k_y\right]} \left(\frac{\Delta^s }{\Delta^t}\frac{\lambda}{4t}
  +\frac{1}{2}\cos k_y+\frac{\mu}{4t} \right) \right.\notag \\
&\quad{} - \left.
  \frac{\nu}{2}\sqrt{\left[\left(\frac{\Delta^s}{\Delta^t}\right)^{\!2}
  -\sin^2 k_y\right]
  \left[1-\left(\sqrt{1-\left[\left(\frac{\Delta^s}{\Delta^t}\right)^{\!2}
  -\sin^2 k_y\right]}+\frac{\Delta^s }{\Delta^t}\frac{\lambda}{2t} +\cos k_y+\frac{\mu}{2t}
  \right)^{\!\!2} \right]}\right\rbrace^{\!1/2}
\end{align}
\end{widetext}
with $\nu= \pm 1$.
The functions $k_m^1(k_y)$ and $k_m^{-1}(k_y)$ describe the upper and lower boundary of $\mathcal{F}_r$, respectively.

\bibliography{Lapp}

\end{document}